\documentclass[twocolumn]{emulateapj}

\newcommand{\Msun}{$M_{\odot}$}

\shorttitle{Disk Evolution in Three Nearby Star-Forming Regions}
\shortauthors{Furlan et al.}
\submitted{To appear in ApJ}

\begin{document}

\title{Disk Evolution in the three Nearby Star-Forming Regions of
Taurus, Chamaeleon, and Ophiuchus}

\author{E. Furlan\altaffilmark{1,2}, Dan M. Watson\altaffilmark{3},
M. K. McClure\altaffilmark{3,4}, P. Manoj\altaffilmark{3}, C. Espaillat\altaffilmark{4},
P. D'Alessio\altaffilmark{5}, N. Calvet\altaffilmark{4}, K. H. Kim\altaffilmark{3}, 
B. A. Sargent\altaffilmark{3}, W. J. Forrest\altaffilmark{3},  L. Hartmann\altaffilmark{4}}

\altaffiltext{1}{NASA Astrobiology Institute, and Department of Physics and 
Astronomy, UCLA, 430 Portola Plaza, Los Angeles, CA 90095}
\altaffiltext{2}{current address: JPL, Caltech, Mail Stop 264-767, 4800 Oak Grove
Drive, Pasadena, CA 91109; Elise.Furlan@jpl.nasa.gov}
\altaffiltext{3}{Department of Physics and Astronomy, University of Rochester, Rochester, 
NY 14627; dmw@pas.rochester.edu, manoj@pas.rochester.edu, khkim@pas.rochester.edu, 
bsargent@pas.rochester.edu, forrest@pas.rochester.edu}
\altaffiltext{4}{Department of Astronomy, The University of Michigan, 500 Church St.,
830 Dennison Bldg., Ann Arbor, MI 48109; melisma@umich.edu, ccespa@umich.edu, 
ncalvet@umich.edu, lhartm@umich.edu}
\altaffiltext{5}{Centro de Radioastronom{\'\i}a y Astrof{\'\i}sica, Universidad Nacional
Aut\'onoma de M\'exico, Apartado Postal 3-72 (Xangari), 58089 Morelia, Michoac\'an, M\'exico; 
p.dalessio@astrosmo.unam.mx}

\begin{abstract}
We analyze samples of {\it Spitzer} Infrared Spectrograph (IRS) spectra of T Tauri
stars in the Ophiuchus, Taurus, and Chamaeleon I star-forming regions, whose median
ages lie in the $<$ 1 to 2 Myr range.
The median mid-infrared spectra of objects in these three regions are similar in shape, 
suggesting, on average, similar disk structures. When normalized to the same stellar 
luminosity, the medians follow each other closely, implying comparable mid-infrared 
excess emission from the circumstellar disks. 
We use the spectral index between 13 and 31 $\mu$m and the equivalent width of the 
10 $\mu$m silicate emission feature to identify objects whose disk configuration departs 
from that of a continuous, optically thick accretion disk. Transitional disks, whose steep 
13-31 $\mu$m spectral slope and near-IR flux deficit reveal inner disk clearing, occur
with about the same frequency of a few percent in all three regions. Objects with 
unusually large 10 $\mu$m equivalent widths are more common (20-30\%); they
could reveal the presence of disk gaps filled with optically thin dust. Based on their
medians and fraction of evolved disks, T Tauri stars in Taurus and Chamaeleon I 
are very alike. Disk evolution sets in early, since already the youngest region, the 
Ophiuchus core (L1688), has more settled disks with larger grains. 
Our results indicate that protoplanetary disks show clear signs of dust evolution at an
age of a few Myr, even as early as $\sim$ 1 Myr, but age is not the only factor 
determining the degree of evolution during the first few million years of a disk's lifetime.
\end{abstract}

\keywords{circumstellar matter --- stars: formation --- stars: pre-main sequence --- 
infrared: stars}

\section{Introduction}

Protoplanetary disks around pre-main-sequence stars are known to dissipate over the
course of a few million years; at an age of $\sim$ 10 Myr, very few objects are still
surrounded by primordial disks \citep{jayawardhana99,haisch01,sicilia06,jayawardhana06,
hernandez06}. 
During the first 1-2 Myr, disks already experience significant evolution: dust grains have 
been shown to grow and settle towards the disk midplane \citep[e.g.,][]{miyake95, 
apai05, dalessio06, furlan05, furlan06}, to be partly processed into crystalline form (e.g., 
\citealt{meeus03, kessler06, sargent06, sargent09}), and to be removed from certain 
radial regions in a disk, likely due to the gravitational interactions with a companion that
formed in the disk, such as a newly formed planet \citep[e.g.,][]{skrutskie90, calvet05, 
espaillat08}. 
By understanding the processes occurring in a protoplanetary disk, we gain insights into 
disk clearing and planet formation mechanisms; observing such disks at the onset of 
these processes allows us to discriminate better the degree to which each process 
contributes to disk evolution.

Close to the Sun, at a distance of about 150 pc \citep{bertout99}, lie three star-forming 
regions whose ages are thought to fall into the 1-2 Myr range: Taurus-Auriga, Ophiuchus, 
and Chamaeleon I. 
Ophiuchus is the youngest ($<$ 1 Myr; \citealt{luhman99}) and most embedded of the 
three; its core region (L1688; $\sim$ 30\arcmin) has a very high stellar concentration 
\citep[e.g.,][]{greene92,bontemps01}. Both Chamaeleon I and Taurus are less 
extinguished and crowded than Ophiuchus. Chamaeleon I has a median age of 2 Myr 
\citep{luhman04a}, but a large age spread ranging from $\lesssim$ 1 Myr up to 6 Myr
\citep{luhman07}, and it is somewhat older than Taurus, whose age is about 1 Myr,
with a distribution of ages between 1 and several Myr \citep{kenyon95,hartmann01,
luhman03}. The region surrounding the compact Ophiuchus core contains several 
young stellar objects (YSOs) that are less obscured and older than the Ophiuchus core 
region \citep{ichikawa89,chen95,wilking05}. Since the location in the H-R diagram of 
objects from this off-core region  \citep[see][]{wilking05} is similar to that of members of 
Chamaeleon I \citep[see][]{luhman07}, these two regions have comparable ages 
($\sim$ 2 Myr), independent of evolutionary models.

The majority of objects in the Ophiuchus, Taurus, and Chamaeleon I star-forming regions
are T Tauri stars, which are initially surrounded by optically thick circumstellar disks that
are accreting onto the star; during this stage, they are referred to as classical T Tauri stars. 
The dust in the disk causes an excess at infrared wavelengths, which decreases over time 
as the disk material is dissipated. The slope $n$ of the spectral energy distribution (SED), 
$\lambda F_{\lambda} \propto {\lambda}^n$, measured between 2 and 
25 $\mu$m, is used to classify systems at different evolutionary stages: Class II objects 
have $-2 < n < 0$ and are surrounded by protoplanetary disks, while Class III objects 
have $-3 < n < -2$ and thus have little or no circumstellar material left \citep{lada87,
adams87,andre94}. Pre-main-sequence stars that do not display any accretion signatures 
are also known as weak-lined T Tauri stars.

The infrared excess from protoplanetary disks is emitted as optically thick, continuum 
emission from the dusty disk interior, and optically thin emission from dust in the disk 
surface layer, whose prominent signatures are silicate emission features at 10 and 
18 $\mu$m. Thus, mid-infrared spectra are ideally suited to study these YSOs; 
they allow us to assess the degree of dust crystallization, grain growth and settling, 
and disk clearing, processes which, at an age of 1-2 Myr, are likely in their initial 
stages. Studying protoplanetary disks in the four regions mentioned above will lead 
to a better understanding of how the circumstellar material evolves and eventually 
dissipates, possibly giving rise to planetary systems like our own.

As part of a large survey of star-forming regions within 500 pc, our team has obtained 
several hundred 5-36 $\mu$m spectra of YSOs in the Taurus, Chamaeleon I, Ophiuchus 
core, and Ophiuchus off-core regions using the Infrared Spectrograph\footnote{The 
IRS was a collaborative venture between Cornell University and Ball Aerospace Corporation 
funded by NASA through the Jet Propulsion Laboratory and the Ames Research Center.} 
\citep[IRS;][]{houck04} on board the {\it Spitzer Space Telescope} \citep{werner04} . 
These spectra are classified and presented in \citet{furlan06} (Taurus), Manoj et al. 
(in preparation) (Chamaeleon I), and McClure et al. (in preparation) (Ophiuchus). 
Here we analyze some two hundred Class II objects identified in those papers.  
In \S\ \ref{obs_data_reduction} we introduce our sample, observations, and data 
reduction and processing, in \S\ \ref{median_sec} we compare the median IRS 
spectra of our objects in Taurus, Chamaeleon I, and the Ophiuchus core region, in 
\S\ \ref{disk_evol_sec} we analyze indicators of disk structure and infer the degree 
of disk evolution in all four regions; finally, we discuss our results in \S\ \ref{discussion} 
and give our conclusions in \S\ \ref{conclusions}.

\section{Observations}
\label{obs_data_reduction}

\subsection{Sample Selection and Data Reduction}

Our Taurus, Chamaeleon I, and Ophiuchus targets were observed as part of an IRS 
guaranteed-time program during IRS campaigns 3, 4, 12, 19, 20, 21, 22, 23, 29, 
and 30, which were scheduled from February 2004 to April 2006. 
The objects\footnote{A table, which lists object names and coordinates, is available 
in the electronic version of the journal; the first few lines of this table are shown in
the Appendix.} were originally selected from infrared catalogs 
of YSOs in these regions that were compiled before the launch of the Spitzer Space 
Telescope in 2003 \citep[mainly the {\it IRAS} Faint Source and Point Source 
Catalogs; the DENIS Database, 2nd release; the 2MASS Catalog; ][]{wilking89, 
greene94, kenyon95, barsony97, cambresy98, persi00}. The targets in the Ophiuchus 
off-core region were chosen from \citet{ichikawa89} and \citet{chen95}; about half 
are located in or close to the L1689 cloud, one object (IRS 60) lies in L1709, and the 
remaining objects are spread over a wider area.
With the availability of the IRS spectra, several targets that previously had an only 
tentative classification could be identified as Class II objects\footnote{Note that 
in Ophiuchus, where extinction is typically high, our Class II sample consists of 
systems that were identified based on the dominance of disk emission and not 
necessarily on their observed 2-25 $\mu$m spectral slope (see McClure et al. 
(in preparation) for details).} (see \citet{furlan06}; McClure et al. (in preparation); 
Manoj et al. (in preparation)). 
In Chamaeleon I, we also included a few very low-mass stars and brown dwarfs 
(spectral type $\gtrsim$ M5) from \citet{luhman04a} and \citet{luhman07}, 
which were observed in IRS campaigns 32, 33, 41, and 42.
Our sample consists of 85 Class II objects in Taurus, 69 in Chamaeleon I,  63 in 
L1688, and 15 in the Ophiuchus off-core regions.

All targets were observed with either the two low-resolution IRS modules 
(Short-Low [SL] and Long-Low [LL], 5.2--14 $\mu$m and 14--38 $\mu$m, 
respectively, $\lambda$/$\Delta\lambda$ $\sim$ 90) or the SL module and 
the two high-resolution modules (Short-High [SH] and Long-High [LH], 10--19 $\mu$m 
and 19--37 $\mu$m, respectively, $\lambda$/$\Delta\lambda$ $\sim$ 600). 
In this way we obtained the full mid-infrared spectrum from 5 to 40 $\mu$m for 
each target.  While in Taurus most objects were observed in mapping mode with
2$\times$3-step maps on the target, in Ophiuchus and Chamaeleon I most objects
were observed in staring mode. The spectra were extracted from the Spitzer Science 
Center's basic calibrated data (BCD) products, either pipeline versions S13.2 or S14.0
(S16.1 for the few objects observed in campaigns 41 and 42), and using the IDL-based
SMART package \citep{higdon04}.
More details on our observing modes, as well as a description of our data reduction steps, 
can be found in \citet{furlan06}; notes on the reduction of individual objects are
mentioned in McClure et al. (in preparation) for Ophiuchus and in Manoj et al. (in
preparation) for Chamaeleon I.

\begin{figure*}
\plottwo{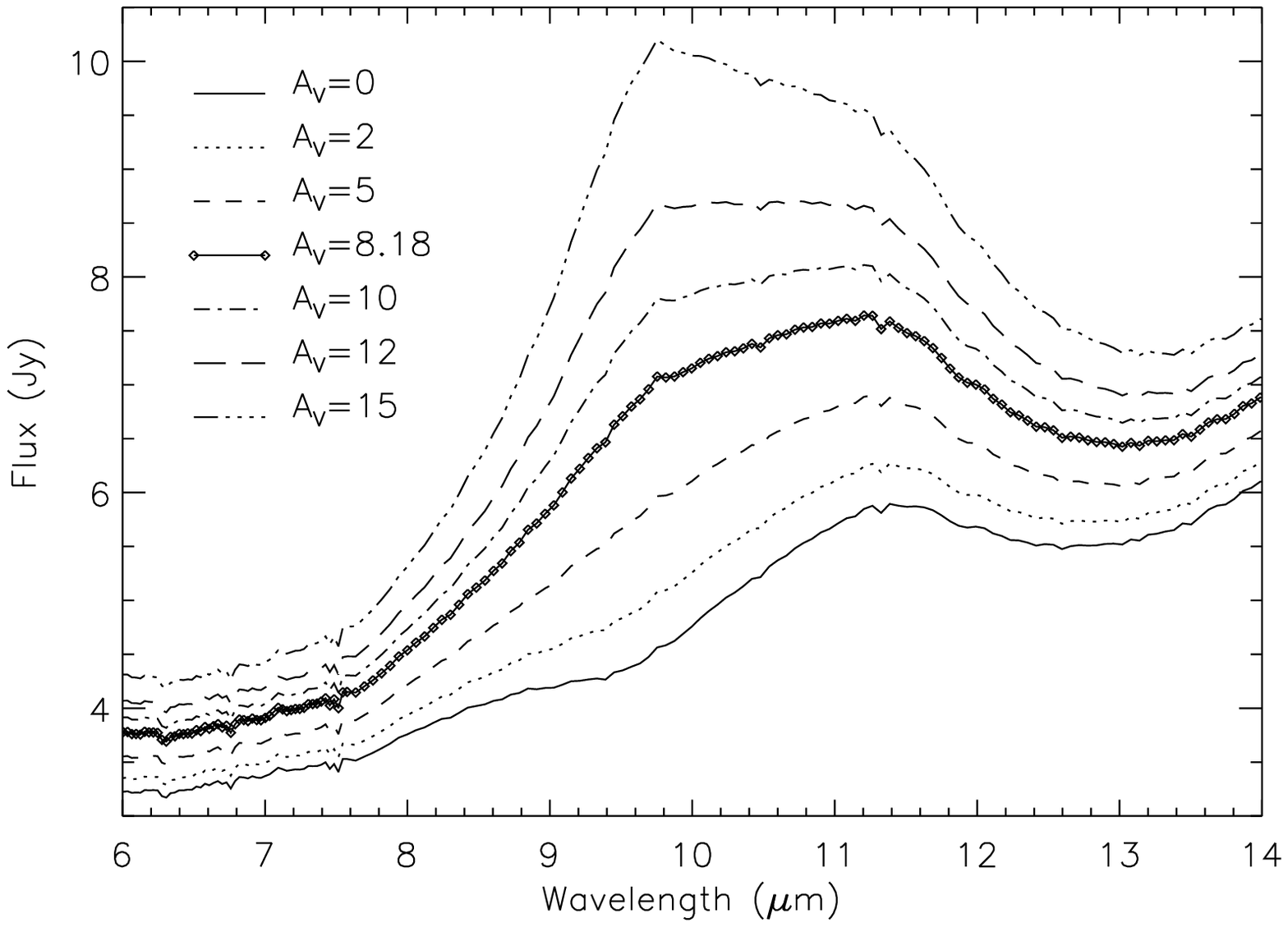}{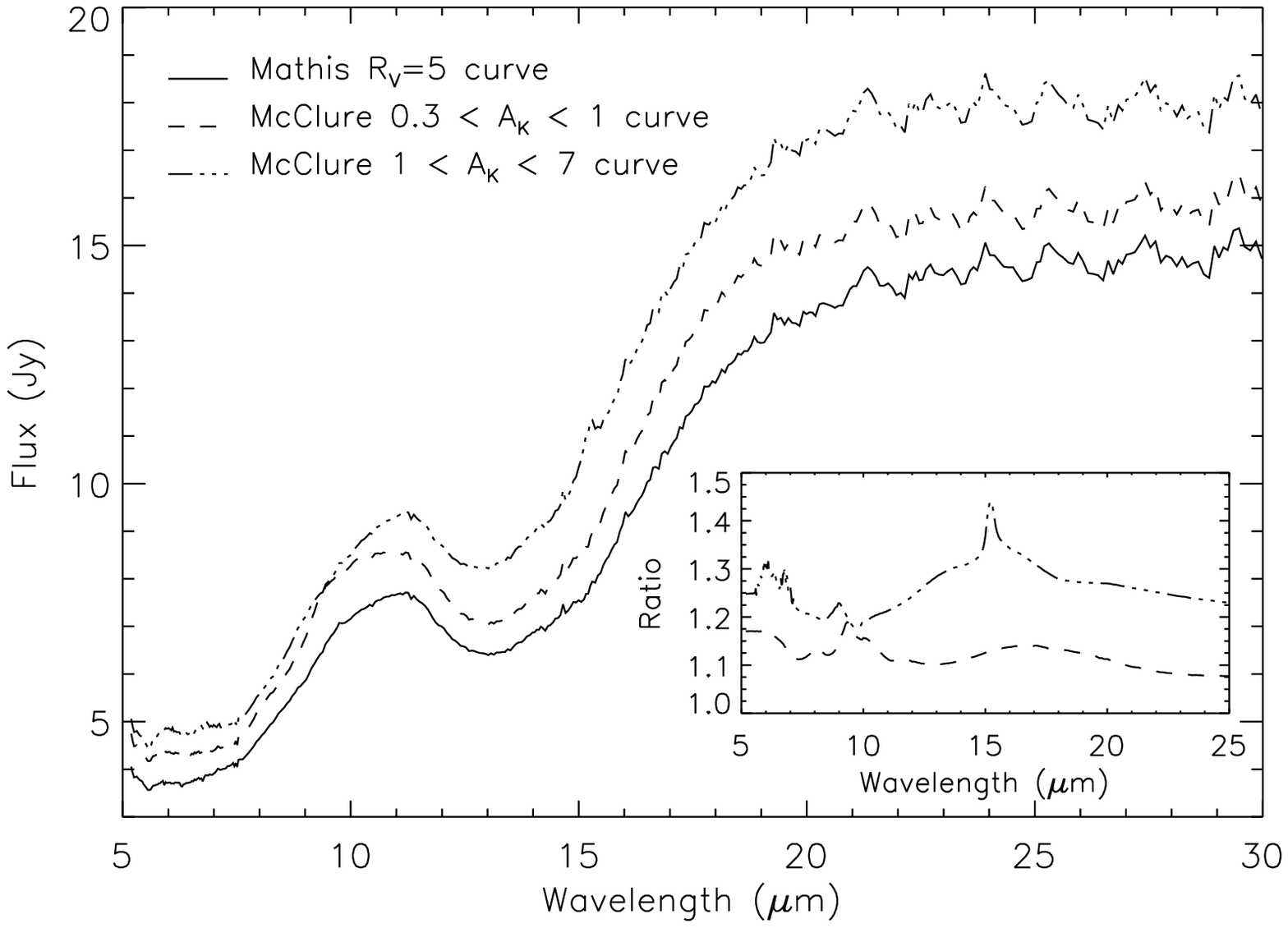}
\caption{Reddening correction for T42, located in Chamaeleon I. {\it Left:} The 
spectrum for $A_V$=0 is the original IRS spectrum, while the other spectra were 
dereddened using Mathis's extinction curve and successively larger $A_V$ values 
(from bottom to top); the actual extinction of T42 is $A_V$=8.18. {\it Right:} 
The IRS spectrum was dereddened assuming $A_V$=8.18 and the three different
extinction curves used in this paper; the spectrum in the middle is the one we
adopted. The inset shows the ratio of the spectra dereddened with the McClure
curves and the one obtained from dereddening with the Mathis's curve. 
\label{Sz32_Av_seq}}
\end{figure*} 

\subsection{Extinction Corrections}

Basic target properties, such as spectral type, extinction, and multiplicity information, 
were taken mainly from the literature 
(Tables \ref{disk_evol_Oph-core}-\ref{disk_evol_Oph-off}). 
In some cases, optical extinction $A_V$ values were not available or resulted in 
dereddened optical photometry inconsistent with the expected photospheric emission. 
For objects with such unknown or uncertain extinction that are located in Taurus or 
Chamaeleon I, we computed the extinction from the observed V-I, I-J, or J-H colors 
by applying Mathis's extinction curve \citep{mathis90} for an $R_V$ value of 3.1. 
For the former two colors, we adopted intrinsic photospheric colors from 
\citet{kenyon95}, while for the latter one, we assumed an intrinsic J-H color typical 
for a classical T Tauri star from \citet{meyer97} (i.e., a typical J-H color defined 
by the CTTS locus). In a few cases we slightly adjusted the derived $A_V$ values 
to obtain agreement between observed and expected photospheric colors.
For objects located in the more deeply embedded Ophiuchus region, McClure et al. 
(in preparation) derived uniform extinctions and uncertainties for the entire sample 
from observed colors (I-J, J-H, or H-K) by applying Mathis's curve for $R_V$=5.0 
(this larger $R_V$ value is more appropriate for dense, molecular clouds; 
\citealt{mathis90}). These authors also obtained near-infrared spectra of most 
of the off-core targets and of several of the more embedded core objects to 
determine spectral types, extinctions, and multiplicity.

In order to create a consistent data set, we decided to adopt the same dereddening 
prescription for all our targets, even though extinction curves likely vary among 
star-forming regions.
For small $A_V$ values, extinction curves for molecular clouds are similar to those 
for the diffuse interstellar medium (ISM), but for $A_V \gtrsim $ 12 there are 
clear deviations \citep{whittet88, chiar07}. Applying an ISM extinction curve like 
the Mathis curve to objects suffering from larger extinctions causes an overcorrection 
of the silicate emission feature at 9.8 $\mu$m, resulting in a feature that is sharply 
peaked at that wavelength, and a 20/10 $\mu$m band ratio that is smaller than 
typical values for amorphous silicates ($\gtrsim$ 0.4; \citealt{dorschner95}).
For example, Figure \ref{Sz32_Av_seq} (left panel) shows the silicate feature of T42, 
a Class II object in Chamaeleon I. The original IRS spectrum displays a depression 
between 9 and 10 $\mu$m caused by extinction along the line of sight; 
when correcting the spectrum for reddening using Mathis's curve, the absorption 
disappears. Using the extinction value determined for T42, $A_V$=8.2, and 
dereddening the spectrum, the silicate emission feature appears more square; 
applying an even larger reddening correction will cause a larger increase around 
9.7 $\mu$m (where the correction applied by the Mathis curve is largest), 
resulting in a sharper silicate feature. 

Studying background objects in molecular clouds, \citet{mcclure09} found that 
already an $A_V$ value of 3 resulted in a mid-infrared extinction curve that was 
different from the Mathis curve; at $A_V$ values larger than about 9, the extinction 
curves converged. \citet{mcclure09} derived two new extinction curves for the 
$A_V \geq$ 3 regime, one for $3 \lesssim A_V < 9 $ ($0.3 \lesssim A_K < 1 $)
and one for $A_V \geq 9$ (the $A_V$ values quoted here assume $R_V$=3.1).
The effect of applying different extinction curves to the IRS spectrum is shown in
Figure \ref{Sz32_Av_seq} (right panel); the McClure curves cause a larger 
correction than Mathis's law, with larger differences at wavelengths where ice
absorptions play a role (see \citet{mcclure09} for details).

We therefore adopted three different extinction curves to deredden our data, which
were applied depending on the $A_V$ value of our targets: we used a spline fit 
to the Mathis $R_V$=5.0 curve for objects with $A_V\,{<}\,$ 3, and the two 
McClure curves for objects with $A_V\,{\geq}\,3$. Since the Mathis and McClure 
laws are given in $A_{\lambda}/A_J$ and $A_{\lambda}/A_K$, respectively, 
we converted the $A_V$ values in Tables \ref{disk_evol_Oph-core} to 
\ref{disk_evol_Oph-off} to $A_J$ and $A_K$ by dividing by 3.55 and 9.0,
respectively. These three curves were also used to deredden fluxes when constructing 
SEDs with ground-based optical photometry from the literature (\S\ \ref{ClassII_n_EW}). 
We note that for IRS spectra the correction for reddening using Mathis's curve for 
$R_V$=3.1 or 5.0 is the same, since the shape of this extinction curve does not 
vary with $R_V$ for near- and mid-infrared wavelengths \citep{cardelli89}.
Furthermore, most objects in Taurus and Chamaeleon I suffer from little extinction 
(the median $A_V$ values for our samples are 1.5 and 2.7, respectively), so the 
details of the extinction curve have little impact on the resulting, dereddened spectra 
and SEDs. On the other hand, the median $A_V$ value for the Ophiuchus core is
13.5, and thus uncertainties in the exact shape of the extinction curve make the
dereddened Ophiuchus spectra somewhat more uncertain.

\subsection{Sample Characteristics}

Our Taurus and Chamaeleon I samples are representative for the Class II population 
of their respective star-forming regions. From a comparison with extensive membership
lists \citep{kenyon95, briceno02,luhman04a}, we conclude that, for spectral types 
earlier than M0, our samples in these two regions are close to complete, while for later 
spectral types (up to M6) we sample at least a few Class II objects in each spectral 
type bin. 

Our sample in Ophiuchus is fairly representative, despite the high extinction in this region
(which renders the determination of spectral types difficult);
the position of our targets in L1688 is coincident with that of the CO gas that defines 
the L1688 cloud \citep{loren89}. While we do not sample the most deeply embedded 
objects in the Ophiuchus core, the median $A_V$ of 13.5 for our L1688 targets suggests 
that several, if not all, of them do belong to the young star-forming core. On the other 
hand, the targets in our Ophiuchus off-core sample (especially those not associated with 
any cloud) are likely members of a distinct, presumably older, population in this region, 
which has recently been surveyed with {\it Spitzer} \citep{padgett08}. Our very 
small sample in the off-core areas implies that our targets there are likely just the 
brightest members with an infrared excess.

\section{Comparison of Median IRS Spectra}
\label{median_sec}

In order to generally assess the state of evolution of circumstellar disks in our
three star-forming regions, we computed the median IRS spectrum for the Class II 
objects in these regions. To avoid including objects with a large range in luminosities, 
we restricted the median to objects with spectral types from K5 to M2 
\citep[see][]{dalessio99}. For Taurus, we used the spectra of 55 Class II objects, 
as in \citet{furlan06}, while for Chamaeleon I and L1688, the spectra of 28 Class II 
objects in each region were included. 

\begin{figure}
\plotone{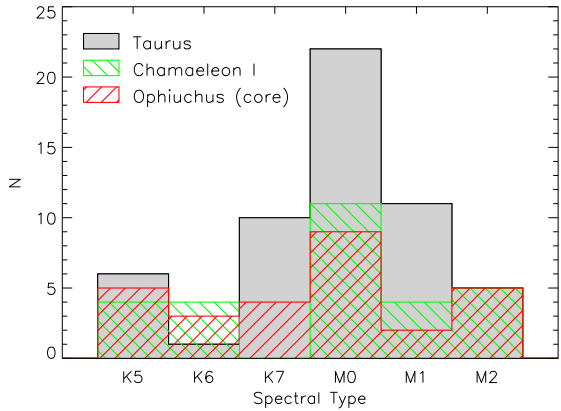}
\caption{Histogram for the distribution of spectral types among the objects entering
the median in the Ophiuchus core, Taurus, and Chamaeleon I star-forming regions.
\label{Tau-Cha-Oph_spec-type_histo}}
\end{figure}

The distribution of spectral types of our median samples is shown in Figure 
\ref{Tau-Cha-Oph_spec-type_histo}. Despite a median spectral type of M0 for all 
three regions, in L1688 43$\pm$12\%\footnote{Note that the uncertainties in the 
number fractions quoted here and in the following sections are based on Poisson statistics 
and thus represent the standard margin of error.}of objects have spectral types earlier 
than M0, while the this fraction amounts to 31$\pm$7\% and 29$\pm$10\% in Taurus 
and Chamaeleon I, respectively. Thus, the median of our Ophiuchus core targets is 
more representative for disks around slightly more massive stars than in Taurus 
and Chamaeleon I.

Before calculating the median for each region, we multiplied the dereddened IRS spectrum
of each object by a scale factor to match the object's dereddened $H$-band flux to the 
median $H$-band flux value of the region (near-IR magnitudes were taken from 2MASS
[\citet{skrutskie06}]). The median IRS spectrum was then computed 
at each wavelength by using the scaled IRS spectra. The normalization at $H$ results in 
an approximate normalization to the stellar luminosities, at the same time minimizing the 
effect of uncertain extinctions \citep[see][]{dalessio99}. 
The L1688 median carries a larger uncertainty due to the often uncertain spectral 
types and visual extinctions of the Ophiuchus sources.

\begin{figure}
\includegraphics[angle=90, scale=0.35]{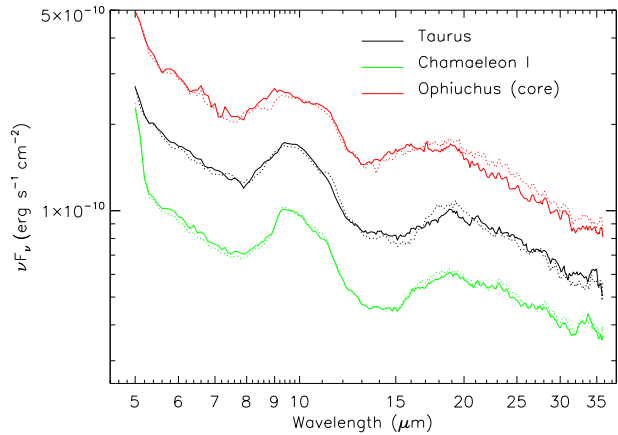}
\caption{The median IRS spectra for Taurus, Chamaeleon I and the Ophiuchus core;
each median spectrum was normalized at a certain, dereddened median flux of the
respective region: median $H$-band flux ({\it solid lines}) and median $J$-band 
flux ({\it dotted lines}). \label{Tau_Cha_Oph_median}}
\end{figure}

In Figure \ref{Tau_Cha_Oph_median} we show the median IRS spectra 
for the Class II objects in Taurus, Chamaeleon I, and the Ophiuchus core region.
In order to check our results, we also computed the medians by normalizing at the
dereddened $J$-band flux, which is more strongly affected by extinction 
corrections, but less by potential near-infrared excess from the circumstellar disks. 
The resulting medians are very similar to the ones normalized at $H$. For the 
remainder of this paper, when we refer to medians, we mean the medians 
that were internally normalized at $H$.
 
\begin{figure}
\includegraphics[angle=90, scale=0.35]{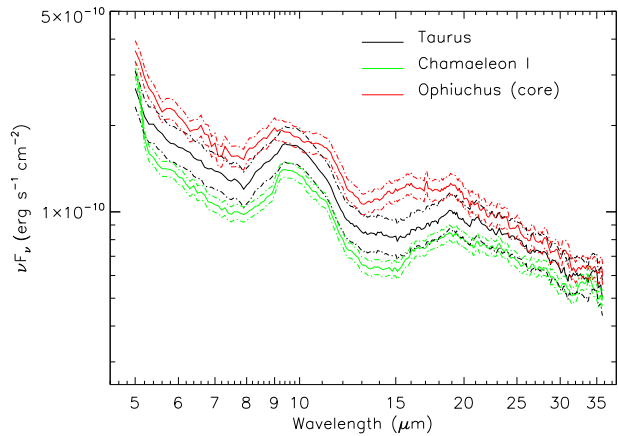}
\caption{The median IRS spectra for Taurus, the Ophiuchus core, and Chamaeleon I,
normalized according to their mean relative distances ({\it solid lines}) and upper and lower 
distance limits ({\it dash-dotted lines}). See text for details.
\label{Tau_Cha_Oph_dist_scaled}}
\end{figure}

The fact that the Chamaeleon I and Ophiuchus medians are about 60\% lower and 
higher, respectively, than the Taurus median can partly be attributed to differences in
the distance to the various regions. Chamaeleon I is at 160-170 pc \citep{whittet97, 
bertout99}, as opposed to $\sim$ 140 pc for Taurus \citep{bertout99, torres07} 
and 120 pc for Ophiuchus \citep{loinard08}. Just due to this diversity, we would 
expect fluxes in Chamaeleon I to be a factor of 0.7 fainter and those in Ophiuchus
to be a factor of 1.4 brighter.
This effect can be seen in Figure \ref{Tau_Cha_Oph_dist_scaled}, where the 
median IRS spectra were multiplied by scale factors according to their relative
distance to Taurus, first by using their mean distance (140 pc for Taurus, 120 pc 
for Ophiuchus, 165 pc for Chamaeleon I), then by considering the distance range 
typically quoted in the literature (130-150 pc for Taurus, 160-170 pc for Chamaeleon I; 
115-125 pc for Ophiuchus; \citealt{bertout99,whittet97,torres07,loinard08}).

After accounting for the larger distance, the Chamaeleon I median approaches the
flux level of the Taurus median, but it is still somewhat lower (on average, at about 
85\% of the Taurus median if the mean distances are adopted). The distance-scaled
median for the Ophiuchus core is still about 20\% higher than the Taurus median.
This could imply different intrinsic $J$- and $H$-band fluxes, with median near-IR
fluxes highest in the Ophiuchus core, followed by Taurus and then Chamaeleon I.

\begin{figure}
\plotone{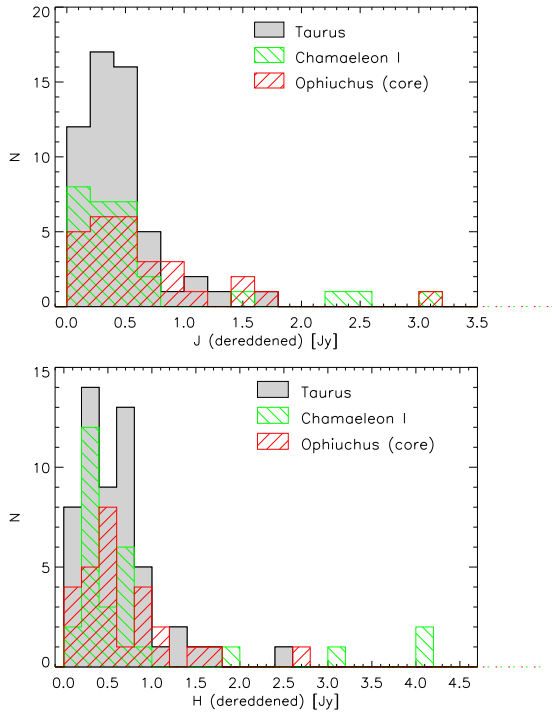}
\caption{Histograms for the distribution of scaled, dereddened $J$- and $H$-band 
fluxes of the objects that entered the median calculation for the Taurus, 
Chamaeleon I, and the Ophiuchus core regions. The flux values were scaled 
according to the mean distance of the various regions with respect to Taurus.
\label{Tau_Cha_Oph_JH_histo}}
\end{figure}

Figure \ref{Tau_Cha_Oph_JH_histo} shows the distribution of dereddened $J$- 
and $H$-band fluxes in the three regions, scaled according to their mean distance 
relative to Taurus to allow a direct comparison of the near-IR flux values. 
The median $J$- and $H$-band fluxes in Chamaeleon I are about a factor of 0.8 
smaller than those in Taurus (the factor decreases to about 0.7-0.75 if the two 
outliers with the highest near-IR fluxes in Chamaeleon I are disregarded); 
therefore stars in Chamaeleon I are intrinsically fainter. On the other hand, the 
median $J$- and $H$-band fluxes of the Ophiuchus core region are higher than 
those of Taurus, by factors of 1.26 and 1.11, respectively (Figure 
\ref{Tau_Cha_Oph_JH_histo}). Thus, the intrinsic near-IR fluxes in L1688 are 
higher than in Taurus, and the higher flux level of the distance-scaled Ophiuchus 
median is likely caused by this difference in near-IR fluxes alone.

\begin{figure}
\includegraphics[angle=90, scale=0.35]{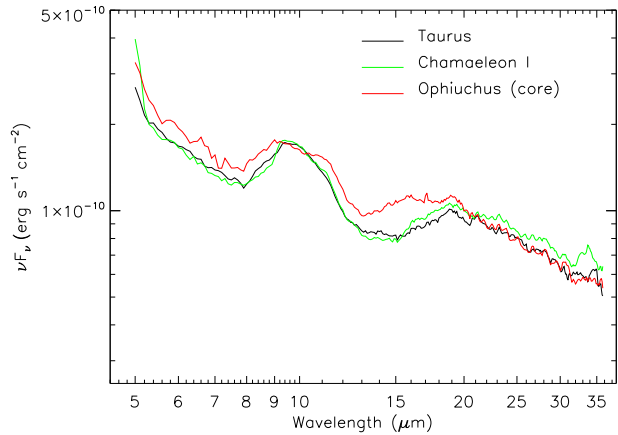}
\caption{The median IRS spectra for Taurus, the Ophiuchus core, and Chamaeleon I,
normalized at the dereddened $H$-band median flux of Taurus. 
\label{Tau_Cha_Oph_Hnorm_median}}
\end{figure}

By normalizing the median IRS spectra of Taurus, the Ophiuchus core, and Chamaeleon I 
to a common $H$-band flux, which amounts to normalization to a common stellar 
luminosity, all three medians essentially overlap (Figure \ref{Tau_Cha_Oph_Hnorm_median}). 
Therefore, the mid-infrared excess emitted by circumstellar disks in these regions is similar 
and proportional to the stellar luminosity; this is expected in typical T Tauri disks, where 
irradiation by the central star dominates the disk heating \citep[e.g.,][]{dalessio99}.
The shapes of the medians are remarkably similar; on a relative scale, the median fluxes 
agree, on average, within a few \%. 
However, in the 5-8 $\mu$m range, the Ophiuchus median is about 15\% higher than 
the Taurus median; this difference almost doubles in the 13-17 $\mu$m region, with a 
maximum deviation of 35\%. This is likely the result of applying the extinction curve for
$A_V \geq 9$ from \citet{mcclure09} to 3/4 of the targets that entered the Ophiuchus
median; this extinction curve exhibits the largest deviation from the Mathis curve and 
causes a larger flux increase around 15 $\mu$m (see the inset in the right panel of 
Figure \ref{Sz32_Av_seq}). Despite some uncertainty in the applicable extinction curve 
for high-extinction regions such as the Ophiuchus core, the small differences we see 
among the medians of the three regions are likely real. 

\begin{figure}
\plotone{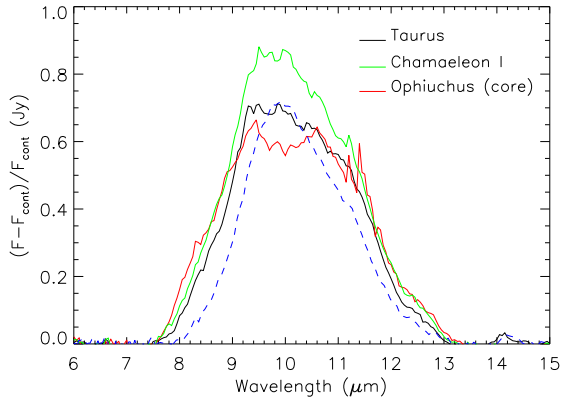}
\caption{The median IRS spectra for Taurus, Chamaeleon I and the Ophiuchus core
around the 10 $\mu$m silicate emission feature, after subtraction and division by a 
continuum fit. The dashed blue line represents the emission feature of small (sub-$\mu$m) 
amorphous silicates; it is the average continuum-subtracted and -normalized profile of 
LkCa 15 and GM Aur, scaled to the mean 9.8 $\mu$m flux of the three medians.
\label{Tau_Cha_Oph_10micron}}
\end{figure}

To compare the 10 $\mu$m emission features, we first fit a continuum, defined as
a 3rd, 4th, or 5th order polynomial, to the 5.6-7.9, 13.0-14.0, 14.5-15.5, 28.0-30.0 
$\mu$m wavelength regions of each spectrum with an associated spectral type of
K5-M2. Then, this continuum fit was subtracted from the spectrum, and the result 
divided by the fit, which results in continuum-subtracted and -normalized 10 $\mu$m 
emission features, i.e., we derived the emission of the optically thin dust alone. The 
median of these 10 $\mu$m features for each region is shown in Figure 
\ref{Tau_Cha_Oph_10micron}. Also displayed in this figure is the average 
continuum-subtracted and -normalized silicate profile of LkCa 15 and GM Aur, 
two of the Taurus objects with the most pristine, or interstellar-like, silicate features 
\citep{sargent09}. 

From Figure \ref{Tau_Cha_Oph_10micron}, it is clear that all three median 10 
$\mu$m features are wider than expected for small (sub-$\mu$m) amorphous 
silicate grains, as are present in the interstellar medium, suggesting that grain 
growth occurred in most disks of all three regions.
The medians of both L1688 and Chamaeleon I display a 10 $\mu$m feature that is 
somewhat more structured than the Taurus median and also wider on the short-wavelength
side. This could be an indication of crystalline silicates, which have narrower and more 
complex emission features than amorphous silicates \citep[e.g.,][]{fabian01,sargent06}; 
in particular, enstatite and silica have strong features around 9 $\mu$m 
\citep{jaeger98, wenrich96}. The fact that the long-wavelength wing of 
the 10 $\mu$m feature is not wider for L1688 and Chamaeleon I as compared to 
Taurus suggests that objects in these two regions do not have, on average, larger grains 
than their counterparts in Taurus. 
These results are just tentative, since only a model of the silicate profiles will reveal the
fractions of large and of crystalline grains, but we believe that the differences we see in
the silicate profiles are real, given that the extinction curves we applied to our spectra
do not introduce artifacts at 10 $\mu$m.

\begin{figure}
\plotone{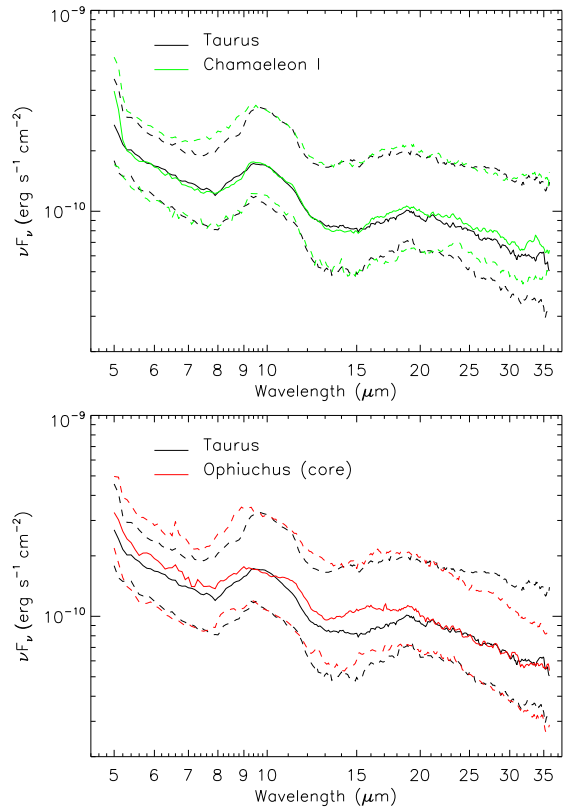}
\caption{The median IRS spectra ({\it solid lines}), normalized at the dereddened 
$H$-band median flux of Taurus, and their quartiles ({\it dashed lines}) for Taurus, 
Chamaeleon I and the Ophichus core. \label{Tau_Cha_Oph_median_quart}}
\end{figure}

Figure \ref{Tau_Cha_Oph_median_quart} shows the median spectra scaled to a
common $H$-band flux (the median of Taurus), together with the quartiles, 
which delineate the range where 50\% of spectra lie. 
For all three regions, the quartiles span a comparable range in flux values, and 
medians and lower quartiles follow each other closely over nearly all wavelengths. 
In the Ophiuchus core region, the lower quartile almost matches that of Taurus,
while the upper quartile is generally a bit higher in flux and decreases more sharply 
beyond about 22 $\mu$m. For Chamaeleon I, both the lower and upper quartiles 
agree very well with the Taurus ones over the 9-22 $\mu$m wavelength region. 
Beyond 22 $\mu$m, the quartiles of the Ophiuchus core and Chamaeleon I define 
a somewhat smaller range in flux values around the median than the Taurus quartiles. 
In Taurus, the lower quartile decrease more steeply than the upper quartile beyond 
about 25 $\mu$m. Overall, the lower quartiles are closer to the medians than the 
upper quartiles, signifying that fainter disks span a smaller range in infrared excess
than the brighter disks.

\section{Comparison of Spectral Indices and Silicate Feature Strength}
\label{disk_evol_sec}

\subsection{Definitions and Models}
\label{n_EW_def_models}

Besides calculating the median mid-infrared spectra, which gauge the typical
distribution of dust in circumstellar disks, we computed two quantities that probe
dust evolution: the spectral index between 13 and 31 $\mu$m, $n_{13-31}$,
and the equivalent width of the 10 $\mu$m silicate emission feature, 
$EW(10\,{\mu}\mathrm{m})$ \citep[see][]{watson09}. 

The former quantity, formally expressed as
\begin{equation}
n_{13-31} = \frac{\log(\lambda_{31}F_{\lambda_{31}})-
\log(\lambda_{13}F_{\lambda_{13}})}
{\log(\lambda_{31})-\log(\lambda_{13})}
\end{equation}
measures the slope of the spectral energy distribution (SED), i.e. $\lambda F_{\lambda}
\propto \lambda^n$. We chose 13 and 31 $\mu$m as the endpoints for our slope
determination, since these wavelengths are dominated by the longer-wavelength continuum 
emission from the optically thick disk. The fluxes at 13 and 31 $\mu$m were 
measured by averaging the flux in the 12.8-14.0 $\mu$m and 30.3-32.0 $\mu$m region,
respectively; these bands exclude the broad silicate emission features at 10 and 20 $\mu$m 
generated in the optically thin disk atmosphere.
As shown in \citet{dalessio06} and \citet{furlan05, furlan06}, a steeper SED is indicative 
of increased dust settling, since a less flared disk intercepts less radiation from the central 
star, resulting in decreased heating and thus fainter continuum emission.

\begin{figure*}
\centering
\includegraphics[angle=-90,scale=0.66]{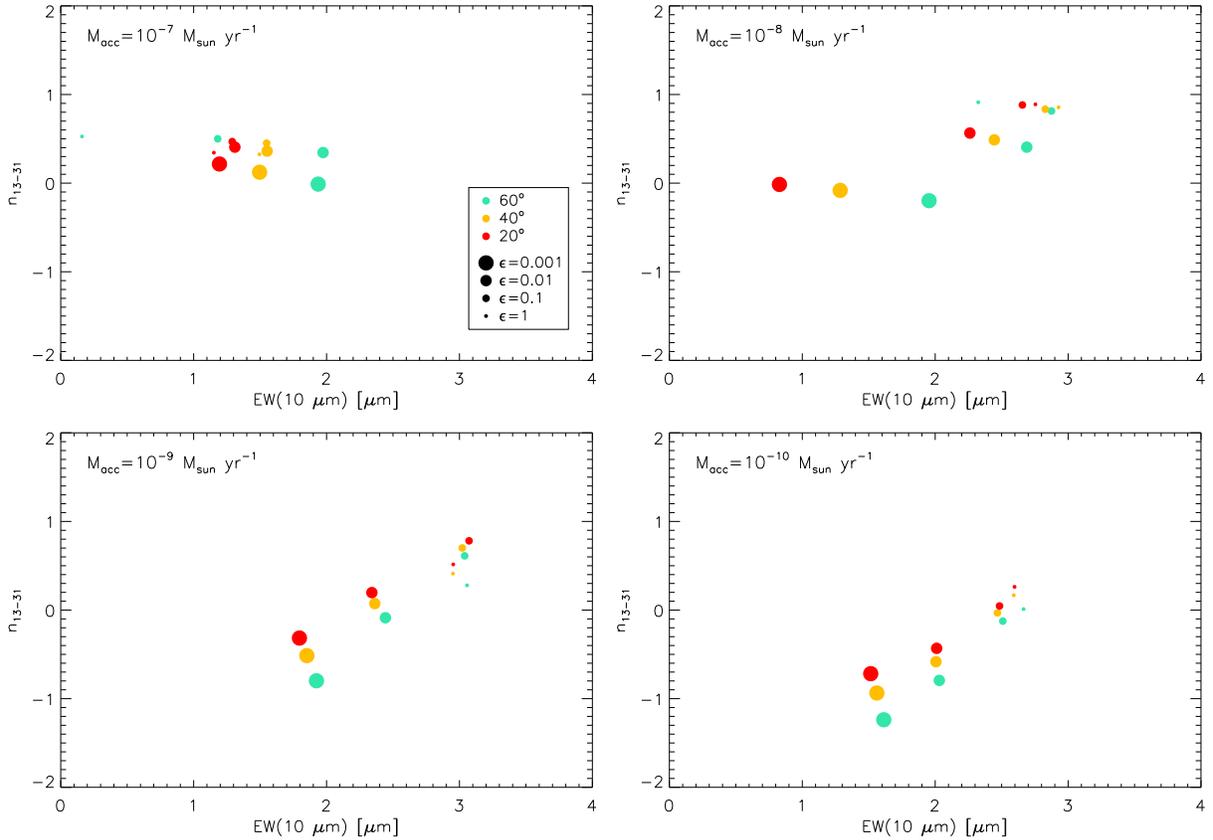}
\caption{Spectral index between 13 and 31 $\mu$m versus the equivalent width of
the 10 $\mu$m silicate emission feature for accretion disk models with mass accretion
rates of 10$^{-7}$, 10$^{-8}$, 10$^{-9}$, and 10$^{-10}$ {\Msun} yr$^{-1}$ 
({\it panels from top left to bottom right, respectively}). The degree of dust settling 
increases according to symbol size. The colors code the inclination angle of the disk, as 
indicated in the label of the first panel. \label{Models_disk_evol}}
\end{figure*}

In order to measure the equivalent width of the 10 $\mu$m feature for each object,
we first defined the continuum as a polynomial anchored in the 5.6-7.9, 13.0-14.0,
14.5-15.5, and 28.0-30.0 $\mu$m wavelength regions. Depending on the object,
we adopted a third-, fourth-, or fifth-order polynomial to represent the continuum
as a smooth curve extending from the wavelength region below 8 $\mu$m to that
beyond 13 $\mu$m. 
Then we subtracted this continuum from the observed spectrum and divided the result 
by the continuum; this normalized, continuum-subtracted spectrum was finally integrated 
from 8 to 13 $\mu$m to yield the equivalent width of the 10 $\mu$m feature:
\begin{equation}
EW(10\,{\mu}\mathrm{m}) = \int_{8{\mu}m}^{13{\mu}m}
{\frac{F_{\lambda}-F_{\lambda,\;cont}}{F_{\lambda,\;cont}}\;d\lambda}
\end{equation}
This definition is actually the exact negative of the standard definition of equivalent
width, which is positive for absorption lines and negative for emission lines. However,
in order to facilitate representations of the equivalent width, we chose the above
definition. The shape of the adopted continuum is often the main source of uncertainty 
of $EW(10\,{\mu}\mathrm{m})$; we took this into account when determining the 
overall uncertainty of our calculated $EW(10\,{\mu}\mathrm{m})$ values.

The $EW(10\,{\mu}\mathrm{m})$ is a measure for the amount of optically thin 
dust per area of optically thick disk. In a typical accretion disk, the optically thin emission 
is generated by small dust grains ($\lesssim$ 5 $\mu$m) in the disk atmosphere 
\citep[e.g.,][]{dalessio99}. It is the emission from these dust grains, and the energy
released by accretion, that heats the lower disk layers; thus, the optically thin emission is 
tied to the optically thick continuum emission from the disk interior, and it will depend on 
the overall structure of the disk. As a disk evolves, dust grains are thought to grow and 
settle towards the disk midplane; the disk becomes less flared (smaller $n_{13-31}$), 
and the strength of the silicate feature decreases 
\citep[e.g.,][]{dullemond04,dullemond05,dalessio06,kessler06,watson09}.

We determined typical ranges for $n_{13-31}$ and $EW(10\,{\mu}\mathrm{m})$ 
by computing these quantities for a grid of accretion disk models around a 0.5 {\Msun}
star, following the same procedures as for the data. We only slightly modified the anchor 
regions for the continuum (7.0-7.9, 13.5-14.5, 28.0-30.0 $\mu$m) due to the different
wavelength grid of the models.
These models were calculated according the methods of \citet{dalessio06} (see also
\citet{espaillat09}); the vertical disk structure was derived self-consistently, resulting in 
a flared disk with a vertical inner rim at the dust sublimation radius (located where the 
dust reaches 1400 K). 
The dust in the disk was assumed to be composed of silicates and graphite \citep{draine84} 
and to scatter isotropically. The degree of dust settling was parameterized with $\epsilon$, 
which is the ratio of the adopted dust-to-gas mass ratio in the upper disk layers relative to 
the standard dust-to-gas mass ratio of 1/100. A small value of $\epsilon$ implies 
depletion of small grains in the upper disk layers and an increase of larger grains close 
to the disk midplane.

The results of our $n_{13-31}$ and $EW(10\,{\mu}\mathrm{m})$ calculations 
for the grid of accretion disk models are shown in Figure \ref{Models_disk_evol}.
For the models with a mass accretion rate of 10$^{-7}$ to 10$^{-8}$ {\Msun} 
yr$^{-1}$, the spread in $n_{13-31}$ is mostly attributed to dust settling, with the 
more settled disks having a steeper SED slope. Dust settling also causes most of the 
spread in $EW(10\,{\mu}\mathrm{m})$ for the 10$^{-8}$ {\Msun} yr$^{-1}$
models, while for the higher mass accretion rate, this spread is mainly caused by 
different inclination angles. We note that at larger inclination angles ($\gtrsim$ 
75\degr; not shown), the silicate feature turns into absorption, especially for the 
more flared disks (i.e., $\epsilon$ $\sim$ 1), resulting in a small and eventually
negative 10 $\mu$m equivalent width. 

We obtain a somewhat smaller spread in $EW(10\,{\mu}\mathrm{m})$ for the 
models with a mass accretion rate of 10$^{-9}$ to 10$^{-10}$ {\Msun} yr$^{-1}$ 
(Figure \ref{Models_disk_evol}), but the range of $n_{13-31}$ values is larger. 
The spread in spectral indices can be attributed to both dust settling and varying 
inclination angles. The more settled, highly inclined model disks approach the spectral 
index of -4/3, which is the value for an infinite, geometrically thin, optically thick disk. 

The less settled model disks have the largest values for $n_{13-31}$ and 
$EW(10\,{\mu}\mathrm{m})$ when their mass accretion rates amount to 
10$^{-8}$ to 10$^{-9}$ {\Msun} yr$^{-1}$, which are representative 
figures for star-forming regions such as Taurus \citep{hartmann98b}. 
In general, when dust grains grow and settle, the $EW(10\,{\mu}\mathrm{m})$ 
decreases; even though the presence of larger grains ($\gtrsim$ 1-2 $\mu$m) 
in the optically thin surface layer would increase the width of the silicate emission 
feature, its amplitude would be reduced such that the $EW(10\,{\mu}\mathrm{m})$ 
still decreases.

\begin{figure}
\plotone{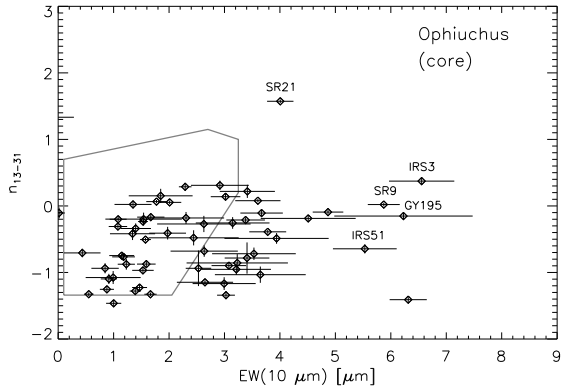}
\caption{Spectral index between 13 and 31 $\mu$m versus the equivalent width of
the 10 $\mu$m silicate emission feature for the Ophiuchus core region. The polygon
specifies the region of typical accretion disks as derived from models. A few of the 
``outliers'', which are discussed in the text, are labeled.
\label{Oph_disk_evol}}
\end{figure}

\begin{figure}
\plotone{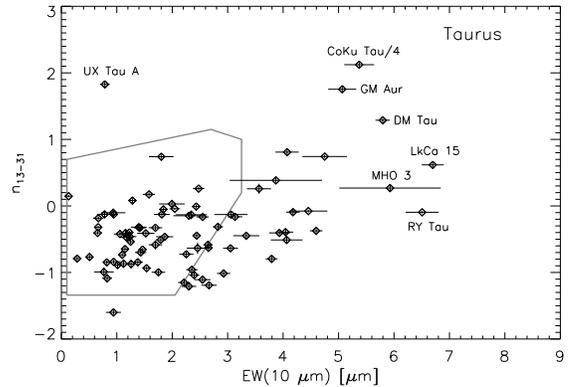}
\caption{Same as Figure \ref{Oph_disk_evol}, but for the Taurus region. 
\label{Tau_disk_evol}}
\end{figure}

\begin{figure}
\plotone{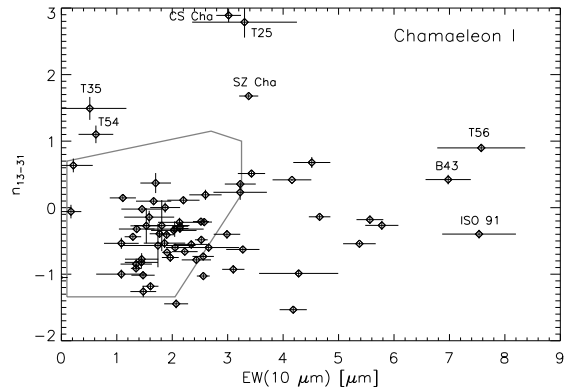}
\caption{Same as Figure \ref{Oph_disk_evol}, but for the Chamaeleon I region. 
\label{Cha_disk_evol}}
\end{figure}

\begin{figure}
\plotone{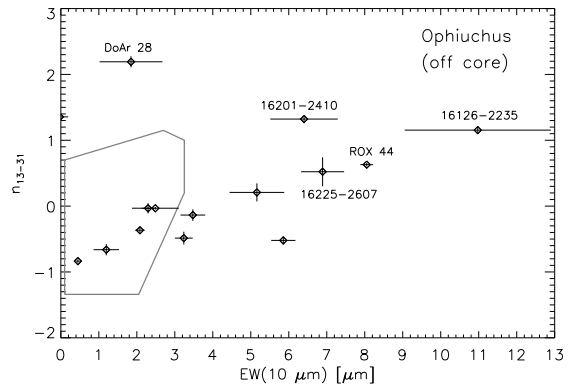}
\caption{Same as Figure \ref{Oph_disk_evol}, but for the Ophiuchus off-core region. 
\label{Oph-off-core_disk_evol}}
\end{figure}

The range of disk models defines a region in the $n_{13-31}$ versus 
$EW(10\,{\mu}\mathrm{m})$ plot which is occupied by typical accretion disks. 
This region extends from -1.33 to about 1.2 in $n_{13-31}$ and from close to 
0 to 2-3 (with higher values for more flared disks) in $EW(10\,{\mu}\mathrm{m})$; 
it is shown as the gray polygon in Figures \ref{Oph_disk_evol} to 
\ref{Oph-off-core_disk_evol}. These latter figures show the spectral index from 
13 to 31 $\mu$m and the equivalent width of the 10 $\mu$m feature calculated 
for the Class II objects in the Ophiuchus core region, Taurus, Chamaeleon I, and 
in the Ophiuchus off-core region, and are discussed below.

\subsection{Class II Objects}
\label{ClassII_n_EW}

\begin{figure}
\plotone{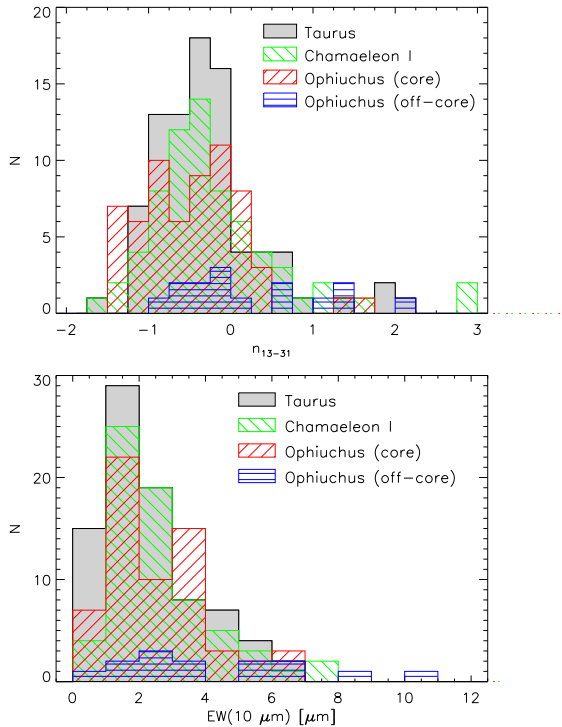}
\caption{Histograms for the distribution of $n_{13-31}$ and 
$EW(10\,{\mu}\mathrm{m})$ values for Taurus, Chamaeleon I, 
the Ophiuchus core and the Ophiuchus off-core regions.
\label{n_EW_hist}}
\end{figure}

As is apparent from Figures \ref{Oph_disk_evol} to \ref{Cha_disk_evol},
most disks in L1688, Taurus, and Chamaeleon I have $n_{13-31}<0$ (see also 
Fig.\ \ref{n_EW_hist}); the fraction is essentially the same in Taurus (80$\pm$10\%)
and L1688 (79$\pm$11\%) and very similar in Chamaeleon I (71$\pm$10\%). 
This indicates that the disk structures of typical objects in these three regions are,
on average, comparable, as was gauged from their medians (\S\ \ref{median_sec}),
and that sedimentation starts early in the lifetime of a protoplanetary disk, 
validating model predictions \citep{goldreich73,chiang97,dullemond05}. 
On the other hand, about half of our small sample of the Ophiuchus off-core 
region, which is similar in age to Chamaeleon I, has $n_{13-31}>0$ 
(Fig.\ \ref{Oph-off-core_disk_evol}). Most of these objects typically have large 
10 $\mu$m equivalent widths, too, suggesting an unusual disk structure.

In fact, each of these regions contains a substantial number of objects with
$EW(10\,{\mu}\mathrm{m})$ values larger than expected from typical 
accretion disks (see Fig.\ \ref{n_EW_hist} for the distribution of 
$EW(10\,{\mu}\mathrm{m})$ values). When counting the number of 
objects outside the region defined by the polygon in Figures \ref{Oph_disk_evol} 
to \ref{Oph-off-core_disk_evol} relative to the total number of objects (not 
including objects whose $EW(10\,{\mu}\mathrm{m})$ is negative due to 
poorly defined continua and silicate features), we derive that 35$\pm$7\%, 
27$\pm$6\%, 29$\pm$6\%, and 60$\pm$20\% of Class II objects in the 
Ophiuchus core, Taurus, Chamaeleon I, and Ophiuchus off-core, respectively, have 
unexpectedly large silicate equivalent widths and/or larger $n_{13-31}$ values. 
Considering the smaller sample size, the Ophiuchus off-core region has a marginally 
larger fraction of unusual objects than the other three regions. There is a lack
of objects with very large $n_{13-31}$ values in the Ophiuchus core region.
We note that some of the uncertainties in $EW(10\,{\mu}\mathrm{m})$
are relatively large; they are the result of ambiguity in the underlying continuum,
which generally affects objects with strong 10 $\mu$m emission more, and
of uncertain extinctions, but the latter to a minor extent and mostly for 
objects in L1688.

\begin{figure*}
\plotone{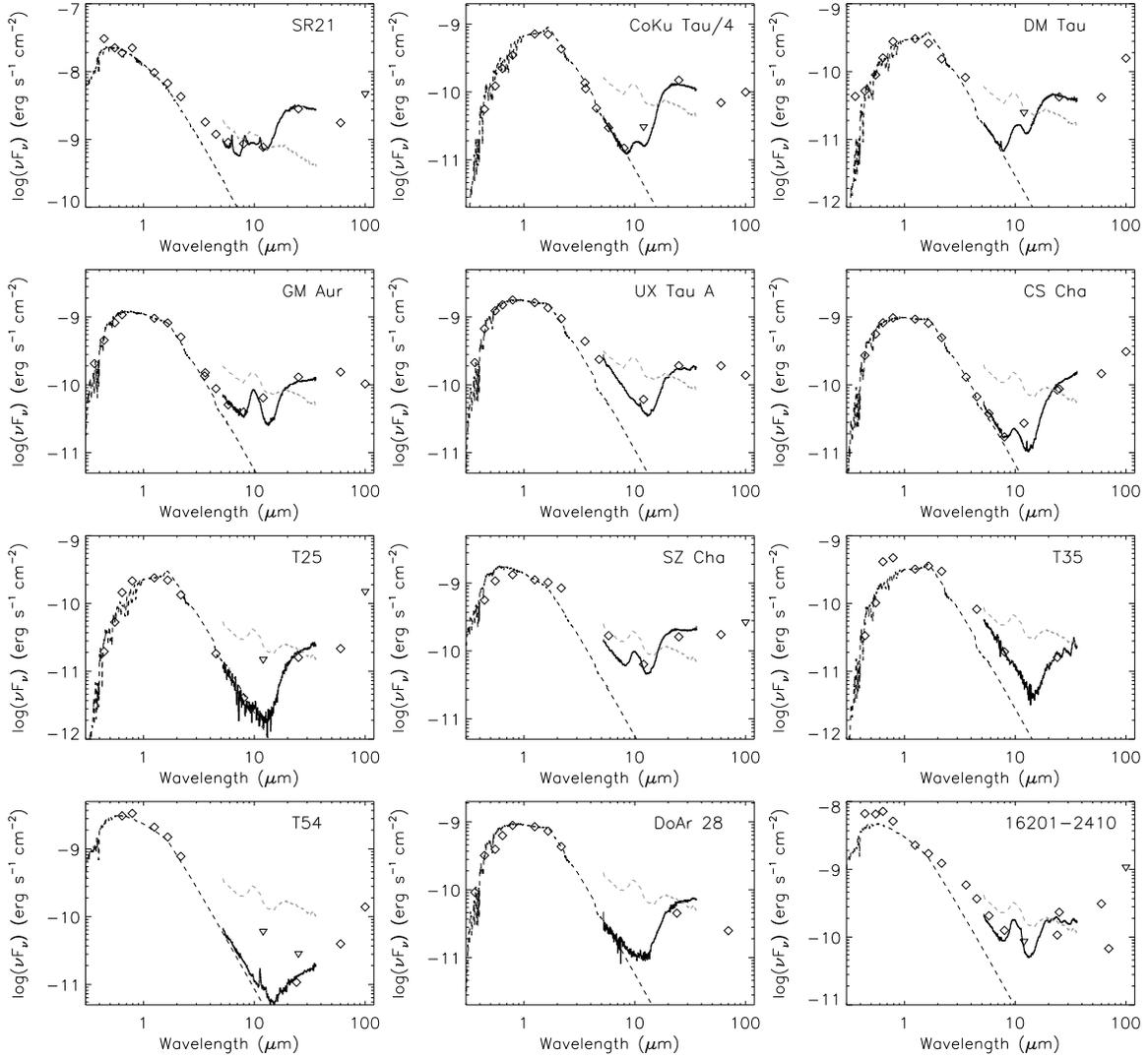}
\caption{The dereddened SEDs of objects with unusually large $n_{13-31}$ values (i.e.,
larger than expected from accretion disk models); the photospheres are represented by
Kurucz model atmospheres with solar metallicity, log(g)=3.5, and different effective temperatures:
T$_{\mathrm{eff}}$=6500 K (SR21), T$_{\mathrm{eff}}$=3625 K (CoKu Tau/4, T25), 
T$_{\mathrm{eff}}$=3750 K (DM Tau), T$_{\mathrm{eff}}$=4750 K (GM Aur), 
T$_{\mathrm{eff}}$=4375 K (UX Tau A, DoAr 28), T$_{\mathrm{eff}}$=4250 K (CS Cha), 
T$_{\mathrm{eff}}$=5250 K (SZ Cha), T$_{\mathrm{eff}}$=3875 K (T35), 
T$_{\mathrm{eff}}$=5500 K (T54), and T$_{\mathrm{eff}}$=6000 K (16201-2410).
The photometry was adopted from the literature, and the data were dereddened as
explained in the text.
The dashed gray lines represent the medians of the Ophiuchus core (for SR21), 
of Taurus (for CoKu Tau/4, DM Tau, GM Aur, UX Tau A), and of Chamaeleon I for the
remaining objects, normalized at the H-band flux of each object. The Chamaeleon I median
was used also for the Ophiuchus off-core targets, since these two regions are of similar age.
\label{n_outliers_SED}}
\end{figure*}

\subsubsection{Outliers in $n_{13-31}$}

We define the objects with steep spectral slopes from 13 to 31 $\mu$m 
($n_{13-31}$ $\geq$ 1) as transitional disks. Despite being Class II objects,
their SED rises sharply beyond about 10 $\mu$m, which is attributed to the 
presence of an inner disk wall bounding an inner disk hole whose deficit in
small dust grains causes a decrease in near-infrared excess emission 
\citep{forrest04,dalessio05,calvet05,espaillat07a}.
A close, stellar companion could be responsible for the disk hole, and a 
luminous, embedded companion, whose SED  peaks in the mid- to far-infrared, 
could create a transitional disk appearance \citep[e.g.,][]{duchene03}.
Therefore, interpretation of transitional disks in terms of disk evolution first 
requires identification of any close companions.

{\it Ophiuchus core.} Only one object associated with L1688 has an unusually large 
$n_{13-31}$ value and reduced near-infrared excess emission: SR21 (see Table 
\ref{disk_evol_Oph-core} and Figure \ref{n_outliers_SED}). Its IRS spectrum rises 
sharply beyond about 14 $\mu$m; in addition, its silicate emission feature is superposed 
with PAH emission features, which are typical for disks around stars of earlier spectral 
type (e.g., \citealt{geers06}; SR 21 has a spectral type of F4), and therefore the 
10 $\mu$m feature strength cannot be determined accurately.
SR21 is a 6.4\arcsec\ binary; the IRS spectrum contains the flux of both components, 
but the secondary is unlikely to contribute significantly at infrared wavelengths 
\citep{prato03}. The near-infrared excess seen in this object rules out a fully 
evacuated inner disk. Since H$\alpha$ measurements indicate that SR21 is not 
accreting any more \citep{martin98}, the inner disk is likely dominated by dust, 
not gas. Models by \citet{brown07} require a gap between 0.45 and 18 AU to 
reproduce the observed SED of SR21, with optically thick disk material remaining in 
the inner disk. Recent high-resolution, sub-millimeter images of this object reveal
an inner cavity of $\sim$ 37 AU in the dust continuum \citep{andrews09},
confirming the presence of the inner disk clearing.

{\it Taurus.} In Taurus, four objects have $n_{13-31}$ values that are well above 
those found in typical accretion disks: CoKu Tau/4, UX Tau A, GM Aur, and DM Tau 
(Table \ref{disk_evol_Tau} and Figure \ref{n_outliers_SED}). 
All have been previously identified as transitional disks with different degrees of
inner disk clearing, and they have been studied in detail by applying SED models
\citep{dalessio05, calvet05,espaillat07b}. The disks of CoKu Tau/4 and DM Tau 
are truncated at 10 and 3 AU, respectively, and the inner regions are depleted in 
dust \citep{dalessio05, calvet05}. GM Aur and UX Tau A have inner disk holes 
at 24 and 56 AU, respectively, but optically thin dust within a few AU (GM Aur) or 
an optically thick ring of material at the dust sublimation radius (UX Tau A) 
\citep{calvet05,espaillat07b}. Recent submillimeter and millimeter interferometer 
maps of GM Aur confirm the inner disk hole of $\sim$ 20 AU \citep{hughes09}. 
Except for CoKu Tau/4, these objects are accreting and do not have close companions 
with mass ratios larger than 0.1 over a 20-160 mas separation range 
\citep{kenyon98,white01,ireland08}. Recent work suggests that CoKu Tau/4 is a 
close binary \citep{ireland08}, and therefore its inner disk clearing is not caused 
by disk evolution. This leaves three ``real'' transitional disks in Taurus.

{\it Chamaeleon I.} In Chamaeleon I, there are five objects with a deficit of 
near-infrared excess emission and large $n_{13-31}$ values: CS Cha, T25, SZ Cha, 
T35, and T54 (Table \ref{disk_evol_Cha} and Figure \ref{n_outliers_SED}; see also 
\citealt{kim09}). 
CS Cha is a transitional disk similar to GM Aur, with some optically thin dust in the inner 
disk region \citep{espaillat07a}. Despite the recent discovery that it is a close binary
system ($\lesssim$ 0.1\arcsec; \citealt{guenther07}), the fact that its inner 
hole is $\sim$ 43 AU in size \citep{espaillat07a} suggests that the binary might 
not be the only agent clearing out the inner disk of CS Cha.
T25 and T54 somewhat resemble CS Cha; their emission is photospheric below about
10 $\mu$m and rises beyond 15 $\mu$m. T25 is a single star and likely not 
accreting \citep{luhman04a,lafreniere08}, implying that disk evolution might be 
responsible for the inner disk hole. T54 is also not accreting \citep{feigelson89}; 
instead of silicate emission, it displays an 11-13 $\mu$m PAH complex, and its 
spectral slope beyond 13 $\mu$m is shallower than that of the other transitional
disks. Since it has a companion at a separation of 0.27\arcsec\ \citep{ghez97}, 
the inner disk is likely cleared by the binary. 
Both SZ Cha and T35 have some near-infrared excess emission and are also still
accreting \citep{gauvin92,luhman04a}, suggesting the presence of optically 
thick material in the inner disk, as is the case for UX Tau A \citep{espaillat07b}. 
While T35 is probably a single star \citep{lafreniere08}, SZ Cha has two ``companions'', 
one at 5.3\arcsec\ and the other at 12.5\arcsec \citep{ghez97}; however, 
these two objects are not confirmed members of Chamaeleon I \citep{luhman07}.
Therefore, we recognize four transitional disks in Chamaeleon I whose inner clearings 
are likely caused by disk evolution.

{\it Ophiuchus off-core.} Finally, we identify two objects in the Ophiuchus off-core 
region with a large 13-31 $\mu$m spectral index and low near-infrared excess 
emission: DoAr 28 and IRAS 16201-2410 (Table \ref{disk_evol_Oph-off} and Figure 
\ref{n_outliers_SED}). The former object lies 0.6\degr\ north of the L1688 cloud 
and is not associated with any of the filamentary clouds in Ophiuchus 
\citep[e.g.,][]{ratzka05}, while the latter one lies about 0.5\degr\ to the west 
of the main L1688 cloud.
The SED of DoAr~28 is reminiscent of that of transitional disks with inner regions devoid 
of dust, like DM Tau or T25, but it might have a slight excess in the 5-8 $\mu$m region. 
Since its H$\alpha$ emission indicates that it is still accreting \citep{cohen79}, the 
inner disk, while depleted in dust, likely contains gas. It does not have a companion 
\citep[the upper limit for the brightness ratio relative to the primary is 0.04 at 
0.15\arcsec;][]{ratzka05}, and therefore its disk structure can probably be seen 
as an indication of disk evolution. 
The IRS spectrum of IRAS 16201-2410 shows a strong silicate emission feature in
addition to the steep rise beyond about 14 $\mu$m. Since, as opposed to DoAr 28,
it has some excess at near-infrared wavelengths, the inner disk is not fully cleared of 
dust. Recent near-infrared observations revealed that it has a companion at a projected 
separation of $\sim$ 2\arcsec\ (McClure et al., in preparation); since this 
companion seems to be very embedded, it could be responsible for a (small) fraction 
of the long-wavelength excess. 

\begin{figure*}
\plotone{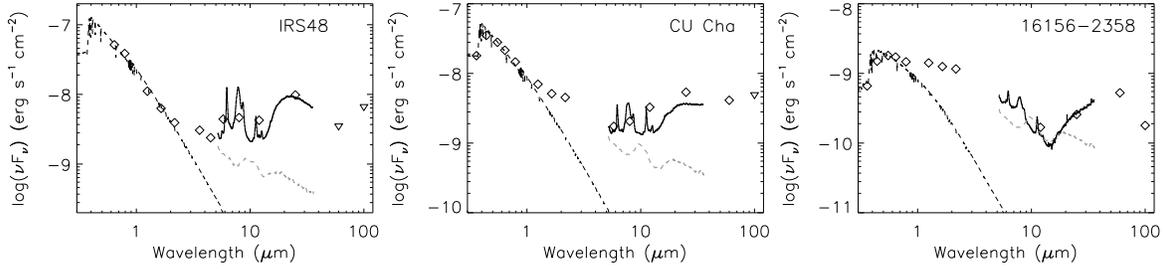}
\caption{The dereddened SEDs of three Herbig Ae/Be stars with large $n_{13-31}$ values; 
the photospheres are represented by Kurucz model atmospheres with solar metallicity, 
log(g)=3.5, and different effective temperatures: T$_{\mathrm{eff}}$=9500 K (IRS48),
T$_{\mathrm{eff}}$=10000 K (CU Cha), and T$_{\mathrm{eff}}$=7250 K (16156-2358). 
The photometry for IRS48 was taken from various published catalogs (USNO, 2MASS, c2d, 
IRAS), that for CU Cha was adopted from \citet{hillenbrand92}, and for 16156-2358 from 
\citet{vieira03}; the data were dereddened as explained in the text.
The dashed gray lines represent the median of Ophiuchus (for IRS48) and of Chamaeleon I
(for the other two objects), normalized at the H-band flux of each object. 
\label{n_AeBe_outliers}}
\end{figure*}

\begin{figure*}
\plotone{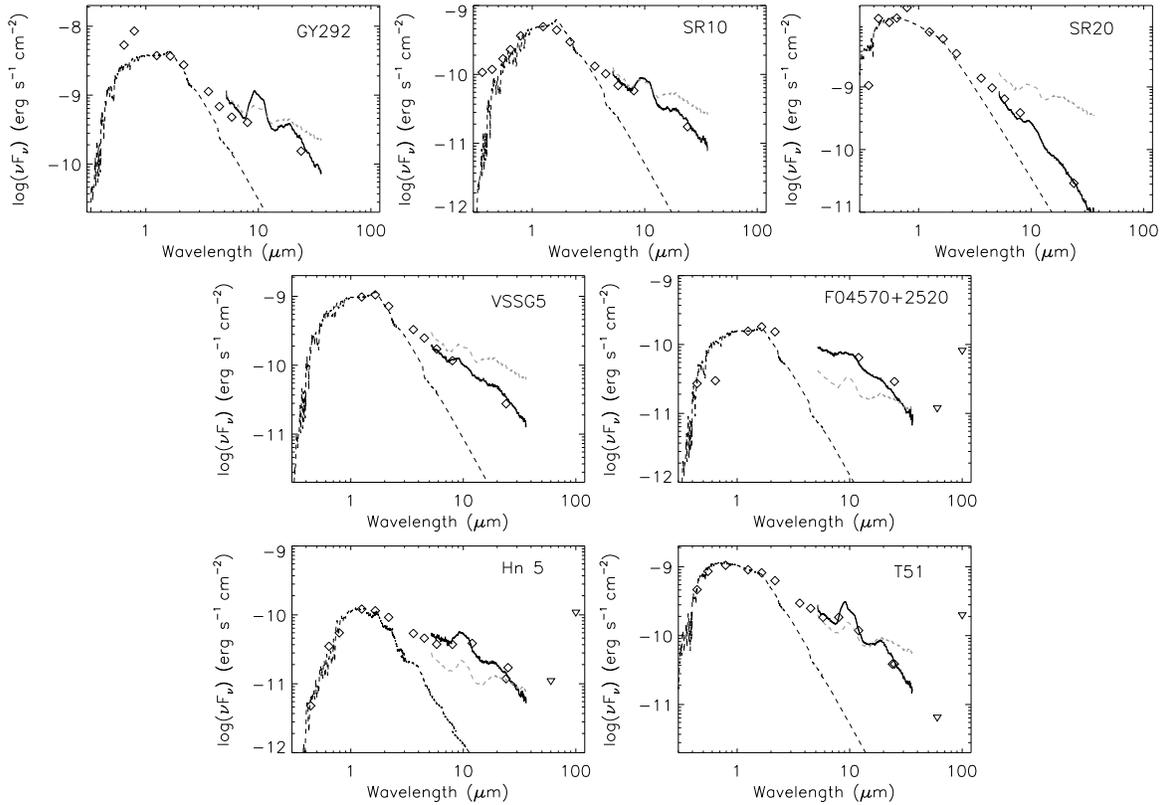}
\caption{The dereddened SEDs of objects with very low $n_{13-31}$ values ($<$ -4/3); 
the photospheres are represented by Kurucz model atmospheres with solar metallicity, 
log(g)=3.5, and different effective temperatures: T$_{\mathrm{eff}}$=4000 K 
(GY292, F04570+2520), T$_{\mathrm{eff}}$=3625 K (SR10), 
T$_{\mathrm{eff}}$=5500 K (SR20), T$_{\mathrm{eff}}$=3875 K (VSSG5), 
and T$_{\mathrm{eff}}$=4750 K (T51). The photosphere for Hn 5 is an AMES-Dusty 
model atmosphere with T$_{\mathrm{eff}}$=3300 K and log(g)=4.0. 
The photometry was adopted from the literature, and the data were dereddened as 
explained in the text.
The dashed gray lines represent the medians of the Ophiuchus core (for GY292, SR10, 
SR20, VSSG5), of Taurus (for F04570+2520), and of Chamaeleon I (for Hn5 and T51), 
normalized at the H-band flux of each object. 
\label{n_low_outliers_SED}}
\end{figure*}

Three more objects have large $n_{13-31}$ values, but display no silicate emission 
feature: IRS48 in L1688, CU Cha in Chamaeleon I, and IRAS 16156-2358 in the 
Ophiuchus off-core (Figure \ref{n_AeBe_outliers}). Their SEDs show numerous PAH 
features and a steep increase beyond about 15 $\mu$m; while CU Cha and IRAS 
16156-2358 have significant near-infrared excess, IRS48 displays no excess below 
2 $\mu$m. These objects are considered Herbig Ae/Be stars, given their early 
spectral type (A0 for IRS48, B9.5 for CU Cha, and F0 for IRAS 16156-2358). 

Models of CU Cha showed that, while the near-infrared excess can be attributed 
to the puffed-up inner rim at the dust destruction radius, the rise in the SED in the 
20-30 $\mu$m region is due to the flaring disk that reemerges from the shadow 
behind the rim \citep{doucet07}. A similar disk structure could apply to IRAS 
16156-2358; in addition, the steep rise of its SED beyond 15 $\mu$m could be 
partly due to a companion 4\arcsec\ to the south, which is fainter than the primary 
at 2.2 $\mu$m \citep[2MASS,][]{skrutskie06}, but whose mid-infrared flux is not 
known. Therefore, despite the large $n_{13-31}$ values, CU Cha and IRAS 
16156-2358 likely do not belong to the transitional disk category.

On the other hand, the lack of near-IR excess emission and the presence of strong 
PAH features in  IRS48 are consistent with observations of \citet{geers07}, who 
resolved a gap with a radius of $\sim$ 30 AU in their 18.7 $\mu$m image of this
object, but found that the PAH emission filled the gap region. Therefore, IRS48 seems 
to have a transitional disk structure, with a lack of small dust grains in the inner disk 
regions, but with some hot dust still present close to the star \citep{geers07}. 

The frequency of disks in a transitional stage, based on $n_{13-31}$ values, is 
comparable in all four regions and amounts to a few \%. It is most uncertain in 
the Ophiuchus off-core region with 13.3$\pm$9.4\%, since the sample size is 
very small. The fraction of transitional disks in L1688 amounts to 3.2$\pm$2.2\%;
we note that L1688 lacks any disks with substantially cleared inner regions that would
result in a severe deficit of infrared excess out to $\sim$~8 $\mu$m. 
The transitional disk fraction in Chamaeleon I (5.8$\pm$2.9\%) is similar to that in 
Taurus (3.5$\pm$2.0\%); the two star-forming environments are also alike.

When examining Figures \ref{Oph_disk_evol} to \ref{Cha_disk_evol} and 
Tables \ref{disk_evol_Oph-core} to \ref{disk_evol_Cha}, a few Class II 
objects with particularly low $n_{13-31}$ values are apparent: GY292, SR10, SR20,
and VSSG5 in the Ophiuchus core, F04570+2520 in Taurus, and Hn 5 and T51 
in Chamaeleon I (see Figure \ref{n_low_outliers_SED} for their SEDs).
The 13-31 $\mu$m spectral index of these objects is actually lower than the 
$-4/3$ value expected for an infinite, geometrically thin, optically thick disk 
\citep[e.g.,][]{adams87}. A possible cause for the steep SED slope is the 
outward truncation of the disk by the gravitational interaction of a close companion 
(\citealt{artymowicz94}; see also \citet{mcclure08} for a model of SR20).
GY292, SR10, F04570+2520, and Hn 5 do not have known companions 
\citep{ratzka05,furlan06,luhman04a}, while SR20 and VSSG5 are sub-arcsecond 
binaries, with separations of 0.04\arcsec-0.07\arcsec\ \citep{ghez95} and 
0.15\arcsec\ \citep{ratzka05}, respectively, and T51 is a $\sim$~2\arcsec\ 
binary \citep{correia06}.

\subsubsection{Outliers in $EW(10\,{\mu}\mathrm{m})$}

While the objects with steep SEDs between 13 and 31 $\mu$m have disks whose
inner regions are depleted in small dust grains to different degrees, the objects with 
large 10 $\mu$m equivalent widths, but more typical $n_{13-31}$ values, are more 
difficult to explain. The optically thin emission is enhanced above levels found for 
accretion disks in hydrostatic equilibrium. Therefore, the ratio of the projected area
of the optically thin medium to that of the optically thick region must be larger than 
in a typical accretion disk. This can be achieved by either increasing the amount or
emitting area of the optically thin dust, or by decreasing the continuum emission 
generated by the optically thick disk regions, but leaving the optically thin emission 
unchanged.

\begin{figure}
\includegraphics[angle=90, scale=0.35]{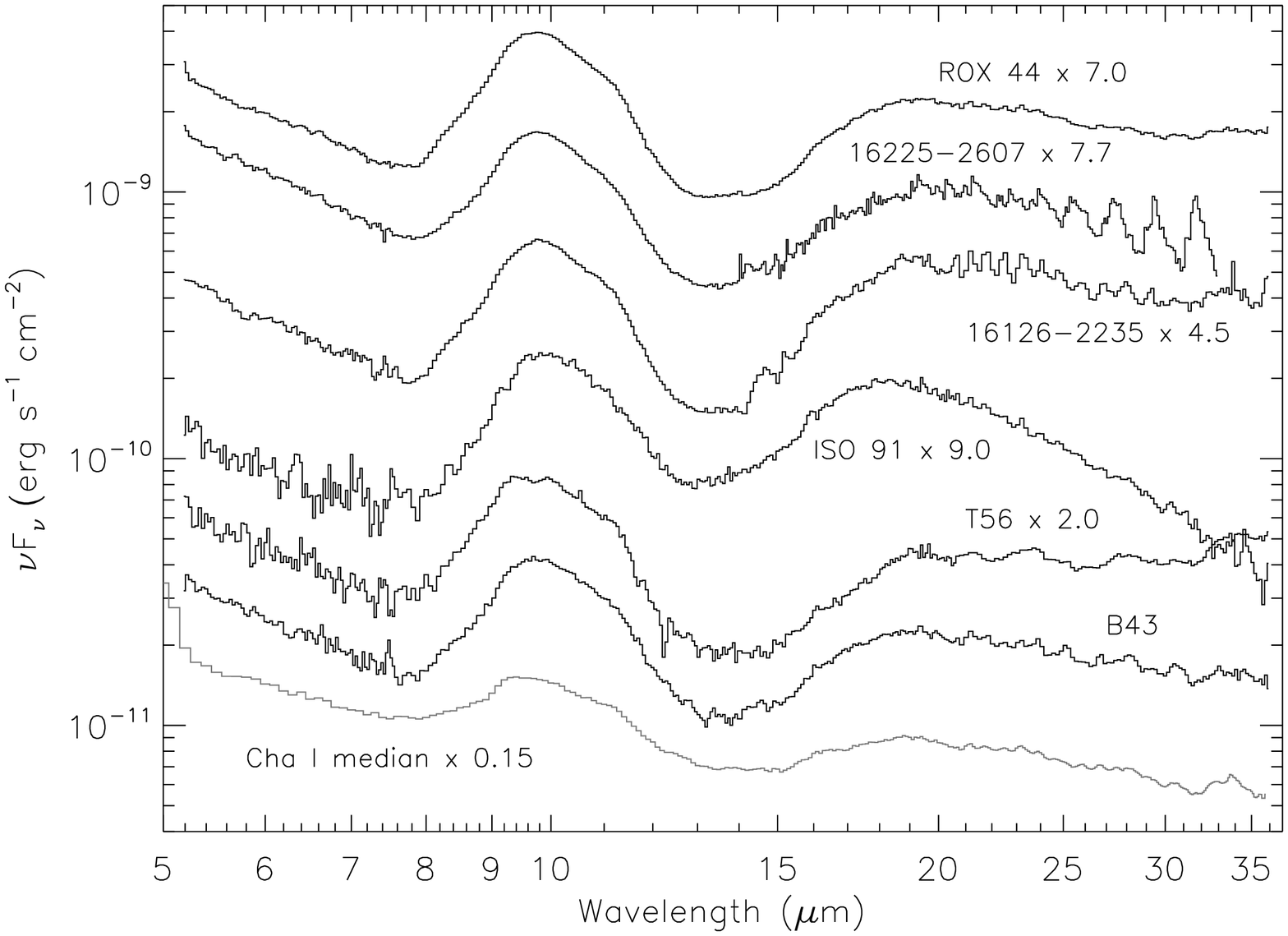}
\caption{The most prominent outliers in terms of $EW(10\,{\mu}\mathrm{m})$
in Chamaeleon I and in the Ophiuchus off-core region, compared to the median of
Chamaeleon I. \label{Cha_Oph-off_outliers}}
\end{figure}

\begin{figure*}
\plotone{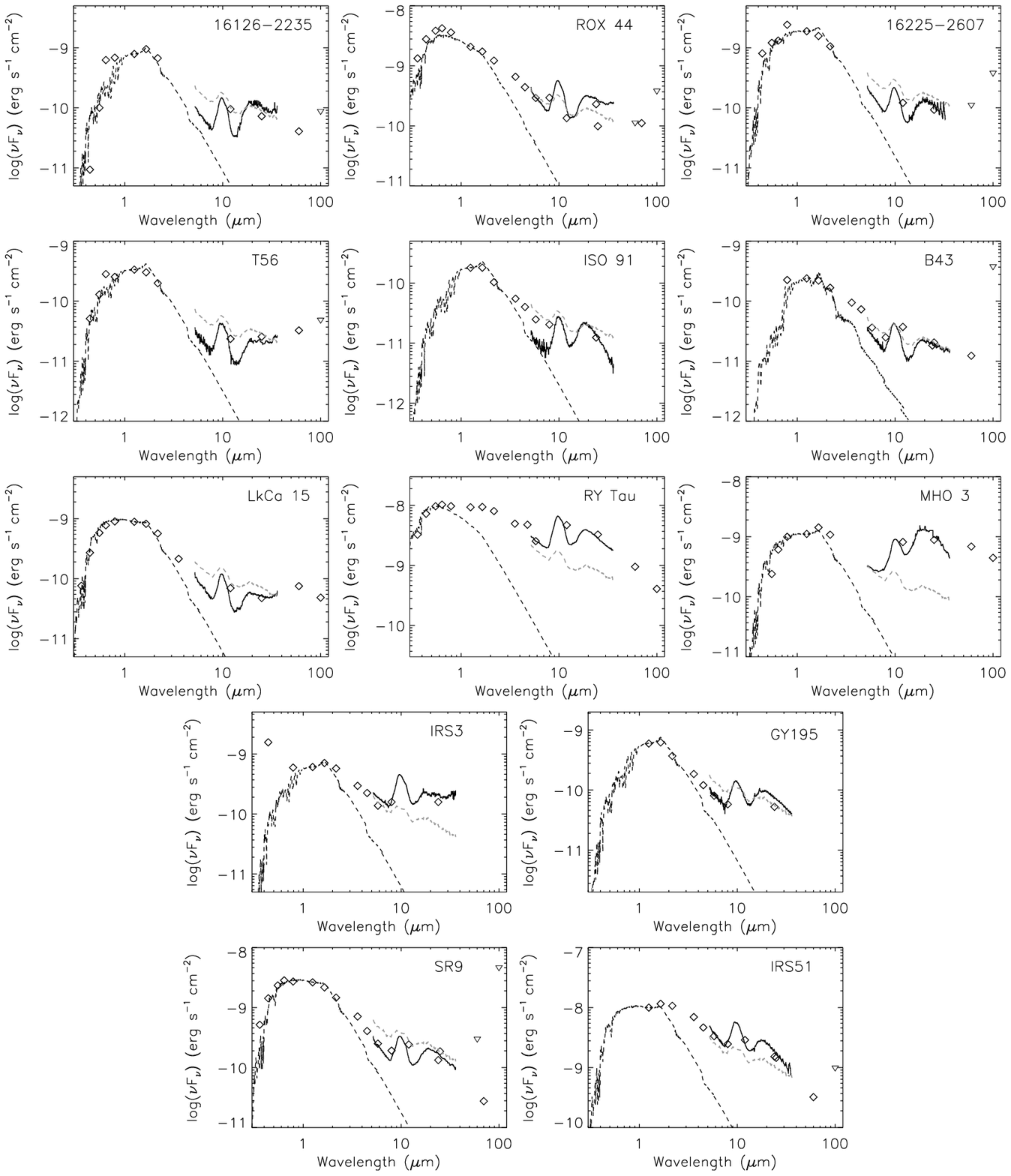}
\caption{The dereddened SEDs of objects with very prominent 10 $\mu$m emission
features; the SED of one additional object that also belongs to this category (GY292) is 
shown in Figure \ref{n_low_outliers_SED}.
The photospheres for all objects except B43 are represented by Kurucz model 
atmospheres with solar metallicity, log(g)=3.5, and different effective temperatures: 
T$_{\mathrm{eff}}$=3500 K (16126-2235, ISO 91, GY195), 
T$_{\mathrm{eff}}$=5250 K (ROX 44),
T$_{\mathrm{eff}}$=4000 K (16225-2607, MHO 3), T$_{\mathrm{eff}}$=3750 K (T56), 
T$_{\mathrm{eff}}$=4375 K (LkCa 15, SR9), T$_{\mathrm{eff}}$= 6000 K (RY Tau),
T$_{\mathrm{eff}}$=3625 K (IRS3), and T$_{\mathrm{eff}}$= 4250 K (IRS51).
The photosphere for B43 is an AMES-Dusty model atmosphere with 
T$_{\mathrm{eff}}$=3500 K and log(g)=3.5. 
The photometry was adopted from the literature, and the data were dereddened as 
explained in the text.
The dashed gray lines represent the medians of Chamaeleon I (16126-2235, ROX 44,
16225-2607, T56, ISO 91, B43), of Taurus (LkCa 15, RY Tau, MHO 3), and of the 
Ophiuchus core (IRS3, GY195, SR9, IRS51), and  normalized at the H-band flux of 
each object.
\label{EW10_outliers_SED}}
\end{figure*}

The most prominent outliers in terms of $EW(10\,{\mu}\mathrm{m})$ are 
found in Chamaeleon I and the Ophiuchus off-core: T56, ISO 91, and B43 in the 
former, and IRAS 16126-2235, 16225-2607, and ROX 44 in the latter (see Figure 
\ref{Cha_Oph-off_outliers}). Of the three off-core targets, only ROX 44 is
associated with a cloud, L1689. Except for IRAS 16126-2235, which is a 1.9\arcsec\
binary \citep{jensen04}, these objects do not have known companions.
Their SEDs display a deficit in near-infrared excess emission, a strong silicate emission 
feature at 10 $\mu$m, and a rising SED beyond 15 $\mu$m, somewhat reminiscent 
of transitional disks (Figure \ref{EW10_outliers_SED}). In fact, their $n_{13-31}$ 
values, while lower than those of transitional disks, lie in a range rarely occupied by 
typical accretion disks. The decreased levels of near-infrared excess emission indicate 
that their inner disks must be depleted in dust, and the rise in the SED at 15 $\mu$m
might suggest the presence of some type of inner disk wall.

\begin{figure}
\includegraphics[angle=90, scale=0.35]{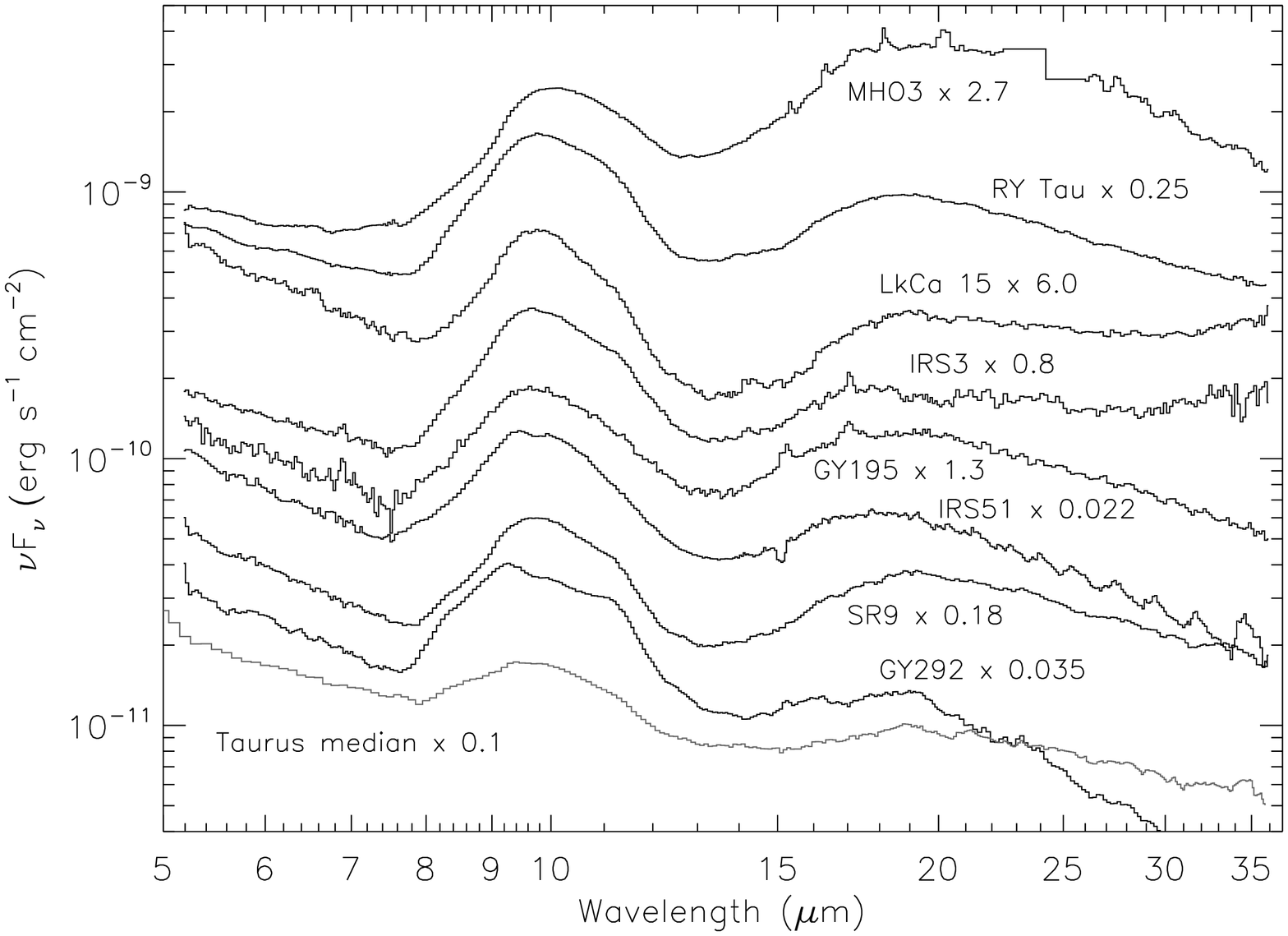}
\caption{The most prominent outliers in terms of $EW(10\,{\mu}\mathrm{m})$
in Taurus and in the Ophiuchus core region, compared to the median of Taurus. 
\label{Tau_Oph_outliers}}
\end{figure}

In the Ophiuchus core region, IRS3, GY195, SR9, IRS 51, and GY292 are notable 
outliers in $EW(10\,{\mu}\mathrm{m})$, while in Taurus LkCa 15, RY Tau, and 
MHO 3 share this characteristic (Figure \ref{Tau_Oph_outliers}). 
While GY195, GY292, SR9 and LkCa 15 have decreased near-infrared excess emission, 
similar to the outliers in Chamaeleon I and the Ophiuchus off-core, IRS3, IRS51, MHO 3,
and RY Tau emit above median levels for their respective star-forming regions over the 
entire IRS wavelength range (Figure \ref{EW10_outliers_SED}). 

IRS3 is a 0.66\arcsec\ binary with a flux ratio of 0.3 at 2.2 $\mu$m \citep{ratzka05}, 
but it is not known to what extent the secondary contributes to the infrared excess.
SR9 also has a companion at a separation of 0.64\arcsec, but its flux ratio of 0.06
relative to the primary at 2.2 $\mu$m \citep{ratzka05} renders it unlikely to significantly
contribute at infrared wavelengths. IRS51 is a binary, too, whose secondary at a distance 
of 1.65\arcsec\ is very faint (flux ratio of 0.04 at 2.2 $\mu$m; \citealt{ratzka05}).
Interestingly, GY292 has both a large $EW(10\,{\mu}\mathrm{m})$ and a 13-31 
$\mu$m spectral index somewhat below -4/3; therefore, its outer disk might be truncated, 
while its inner disk shares some of the properties of the other 
$EW(10\,{\mu}\mathrm{m})$ outliers.

RY Tau is the only object among the sample of  $EW(10\,{\mu}\mathrm{m})$ 
outliers with an early spectral type (G1; see Table \ref{disk_evol_Tau}). It might have 
a very close companion, at a separation of at least 24 mas \citep{bertout99}. Based 
on a dip in its mid-infrared SED, \citet{marsh92} inferred that a gap could be present 
in its disk. On the other hand, recent models suggest that the SED and mid-infrared 
visibilities of RY Tau can be reproduced with an accretion disk surrounded by an optically 
thin envelope, possibly replenished by a disk wind \citep{schegerer08}.

LkCa 15, which is likely a single star \citep{leinert93}, has been recently modeled as a 
disk with a $\sim$~46 AU inner hole that is not fully cleared out. An optically thick inner 
disk wall at the dust sublimation radius and an optically thin inner region extending out 
to a few AU were required to reproduce the near-infrared excess and 10 $\mu$m 
silicate emission \citep{espaillat07b,espaillat08}. High-resolution interferometer images 
at millimeter wavelengths resolved the inner $\sim$ 50 AU hole of LkCa 15 in the dust 
continuum \citep{pietu06}, validating the SED model. It is possible that the other 
objects with an unusually large $EW(10\,{\mu}\mathrm{m})$, decreased near-
and mid-infrared excess, and steep SED rise beyond $\sim$ 15 $\mu$m have 
such a disk gap filled with small, optically thin dust.
In fact, \citet{andrews09} recently imaged an inner hole of $\sim$ 33 AU in the 
disk of ROX 44 using high-resolution sub-millimeter observations; they also note that,
given that this object is still accreting and has a notable infrared excess, material
must be present within the cavity detected in the millimeter continuum emission. 

The fraction of objects with unusually large $EW(10\,{\mu}\mathrm{m})$, but 
typical $n_{13-31}$ values, amounts to 30$\pm$7\%, 21$\pm$5\%, 
19$\pm$5\%, and 47$\pm$18\% in the Ophiuchus core, Taurus, Chamaeleon I, 
and Ophiuchus off-core region, respectively. Thus, Ophiuchus has a marginally larger 
fraction of objects with an uncommon disk structure. Even though the Ophiuchus core 
suffers from larger extinction as compared to the other regions, we believe that the 
use of the McClure extinction curves for $A_V\,{\geq}\,3$ does not overcorrect 
the 10 $\mu$m silicate emission feature, and therefore our derived values of 
$EW(10\,{\mu}\mathrm{m})$ are valid. Also, of the seven Ophiuchus off-core 
outliers, only two have $A_V\,{\geq}\,9$, and the four objects with the largest 
$EW(10\,{\mu}\mathrm{m})$ values suffer from relatively little extinction.

\section{Discussion}
\label{discussion}

\subsection{First Steps of Disk Evolution}

Circumstellar disks in the nearby star-forming regions of Ophiuchus, Taurus, and 
Chamaeleon I display clear signs of disk evolution. When considering the spectral
index between 13 and 31 $\mu$m and the equivalent width of the 10 $\mu$m 
silicate emission feature, outliers that do not fall in the region expected for typical 
accretion disks reveal altered, likely more evolved, disk structures. 
However, most of the objects have $n_{13-31}$ and $EW(10\,{\mu}\mathrm{m})$
values that are not unusual (see Figures \ref{Oph_disk_evol} to 
\ref{Oph-off-core_disk_evol}). When comparing them to model calculations, the fact 
that most disks have a negative spectral index suggests dust depletion factors of 
10$^2$ to 10$^3$ in the upper disk layers, as was already found for Taurus 
\citep{furlan06}. Hardly any objects occupy the region where accretion disks with 
typical mass accretion rates of 10$^{-9}$ to 10$^{-8}$ {\Msun} yr$^{-1}$ and 
$\epsilon=1$ lie, even in the young Ophiuchus core, implying rapid dust growth and 
settling already at an age of $\lesssim$ 1 Myr.

In previous work \citep[e.g.,][]{bouwman01,vanboekel03,meeus03,przygodda03}
the strength of the 10$\mu$m silicate feature was analyzed in a different way:
it was estimated from the peak-over-continuum ratio in the 10 $\mu$m region,
and it was compared to the flux ratio of the normalized spectrum at 11.3 and 9.8 
$\mu$m, taken as a measure for the shape of the silicate feature. For both 
Herbig Ae/Be stars and T Tauri stars, a correlation between shape and strength
is observed \citep[e.g.,][]{vanboekel03,kessler06}; strong features typically have low 
F$_{11.3}$/F$_{9.8}$ ratios, which is interpreted as emission from small, 
amorphous silicate grains, as are found in the interstellar medium. Conversely, 
a large F$_{11.3}$/F$_{9.8}$ ratio (and weak silicate feature strength) 
suggests the presence of larger and/or more crystalline grains, i.e., more 
processed dust. 

Therefore, the large 10 $\mu$m equivalent widths we observe in some of our
spectra correspond to large peak-over-continuum ratios; they could be explained 
as objects with particularly large mass fractions of small grains in their disk surface
layers. However, this interpretation does not take the optically thick disk component 
into account; if there were more small grains in the disk atmosphere absorbing stellar 
radiation, the lower disk layers would be heated more, resulting in an increase of 
continuum emission and no substantial change in the 10 $\mu$m equivalent width.
We checked this assumption by calculating new accretion disk models, which differ
from the ones presented in \S\ \ref{n_EW_def_models} only by the maximum
grain size in the upper disk layers, 0.05 $\mu$m as opposed to 0.25 $\mu$m
in the previous models. For these new models, the 10 $\mu$m equivalent width 
increased by at most 0.2 $\mu$m, proving that large $EW(10\,{\mu}\mathrm{m})$ 
values cannot be explained by a large abundance of very small grains in the disk 
atmosphere. Also changing the dust composition does not significantly increase 
$EW(10\,{\mu}\mathrm{m})$; we calculated models using opacities for
amorphous silicates of olivine composition from \citet{dorschner95}, which
resulted in $EW(10\,{\mu}\mathrm{m})$ values of up to 4 $\mu$m for
the less settled models \citep{espaillat09}.

As noted by \citet{watson09}, the interpretation of peak-over-continuum as 
different degrees of grain processing neglects the likely diversity in the structure of 
the underlying optically thick disk due to dust settling. A more settled disk will have 
decreased continuum emission and typically also less emission from the optically thin 
disk surface layers. Thus, our $EW(10\,{\mu}\mathrm{m})$ values and also 
the peak-over-continuum ratios do not necessarily probe different grain sizes in the 
disk atmosphere, but rather are a measure for disk structure in terms of the ratio 
of the optically thin and optically thick emission.

\subsection{Formation of Disk Structure}

The spectral index measures the degree of dust settling in a disk, but is also 
sensitive to disk inclination, especially for low mass accretion rates ($\lesssim$
10$^{-9}$ {\Msun} yr$^{-1}$). An $n_{13-31}$ value larger than expected 
for a disk with no dust settling (i.e., $\epsilon$=1) is indicative of a substantial
change in the structure of the inner disk; all transitional disks studied so far 
in the Taurus and Chamaeleon I regions have large $n_{13-31}$ values 
and inner disk clearings ranging from a few to several tens of AU
\citep{dalessio05,calvet05,espaillat07a,espaillat07b,kim09}. 

Close binary companions create inner disk gaps whose size depends on the 
orbital parameters of the binary \citep{artymowicz94}. Such systems would 
appear as transitional disks, but in fact be long-lived circumbinary disks. The
companion star fraction in the young star-forming regions of our study is fairly
large, $\sim$~50\% (for projected separations in the 15-1800 AU range; 
\citealt{ghez97}), but it decreases for smaller separations of the binary 
components ($\sim$ 5\% for 0.02-0.5 AU projected separations;
\citealt{melo03}). More specifically, 40--50\% of the stars in our sample are 
found in multiple systems, but not all objects have been observed at sufficiently
high resolution to determine the presence of companions within $\sim$ 1\arcsec.
Thus, it is likely that a few of the transitional disks are the result of clearing by a 
central binary. In fact, CoKu Tau/4 and CS Cha have recently been found to be 
close binary systems \citep{ireland08,guenther07}, but it seems that only in 
CoKu Tau/4 are the stars responsible for inwardly truncating the disk.

On the other hand, disks are thought to evolve from inside out; processes
occur on shorter timescales in the inner disk, where densities are higher and 
orbital speeds larger \citep[e.g.,][]{dullemond04,alexander07}.
Substantial dust growth, planet formation, photoevaporation, or draining of the 
inner disk induced by the magneto-rotational instability (MRI) results in inner 
disk clearing \citep{marsh92,clarke01,dalessio05,alexander07, chiang07, kim09}. 
If caused by these mechanisms of disk evolution, transitional disks are in a 
temporary state, since processes driving the clearing of the inner disk will 
eventually engulf the whole disk. The fact that we see evidence of inner disk 
clearing at the young age of L1688 ($<$ 1 Myr) indicates that disk evolution 
starts early. 

The frequency of transitional disks in the three main regions we studied (3--6\% 
of objects at an age of 1-2 Myr) implies a timescale of several 10$^4$ up to 
$\sim$ 10$^5$ years for the transitional disk phase, assuming disk evolution
is continuous and that all disks experience this stage in their evolution. It is also
possible, but unlikely, that only 3--6\% of disks in a star-forming region appear 
as transitional, with the mechanism that causes the inner disk hole active for a 
long period ($\sim$ 1-2 Myr) of time, such as in a circumbinary disk. Disks are
known to dissipate over time, with clearing processes on the timescale of $\sim$
10$^5$ years, as model calculations for disk clearing by planet formation 
\citep[e.g.,][]{quillen04,varniere06}, photoevaporation 
\citep[e.g.][]{alexander06}, and MRI-activated disk draining \citep{chiang07}
have shown.
The lifetime of a transitional disk is thus shorter than that of a primordial disk 
($\sim$ a few Myr). Recent work by \citet{currie09} suggests a long 
transitional disk timescale, but this is mainly due to the different definition of 
transitional disks; those authors consider all ``evolved'' disks as transitional, 
while we only include disks that show evidence of inner disk clearings. By 
adopting a similar definition of transitional disks as we do and using high-quality 
{\it Spitzer} data for Taurus and a consistent method of estimating disk 
fractions, \citet{luhman09} show that transition timescales appear to be 
considerably less than primordial disk lifetimes, substantiating our result.

The equivalent width of the 10 $\mu$m silicate emission feature could also suggest 
changes in the configuration of the disk, especially when accompanied by a decrease
in the near- and mid-infrared emission. A large $EW(10\,{\mu}\mathrm{m})$ 
implies stronger emission by optically thin dust relative to the optically thick disk emission; 
a gap in the disk, filled with optically thin dust, could create this signature and would 
be a precursor of a transitional disk with an inner disk hole \citep{espaillat07b,espaillat08}. 
On the other hand, an object with a dust-free gap would not exhibit a large 
$EW(10\,{\mu}\mathrm{m})$, as is the case for UX Tau A \citep{espaillat07b}.
An alternative explanation for the $EW(10\,{\mu}\mathrm{m})$ outliers, as shown 
by modeling of RY Tau \citep{schegerer08}, is the presence of an envelope of optically 
thin dust that could be a source for the enhanced optically thin emission seen in these 
objects. However, this envelope would have to be replenished in small dust grains by 
some type of flows originating in the disk, and it is not known whether sufficiently high 
mass loss rates could be achieved.

Assuming most objects with large $EW(10\,{\mu}\mathrm{m})$ have developed
gaps in their disks and that all disks experience this evolutionary phase once in their
lifetime, the larger number of objects presumably with gaps in their disks compared 
to that of objects with inner holes ($\sim$ 20--30\% of objects at an age of 1-2 Myr) 
implies that the stage in which a disk has a gap lasts longer than the period of inner disk 
clearing, from several 10$^5$ up to 10$^6$ years. Alternatively, a disk with a gap 
could survive for a shorter period of time, but then experience this phase a few times 
before being dissipated.
In addition, even if just a few of the disks with large $EW(10\,{\mu}\mathrm{m})$ 
in the Ophiuchus core were disks with gaps, the presence of such disks in this
young region would suggest that gap formation sets in early. It is interesting that
$\sim$ 50\% of the objects in the Ophiuchus off-core region have unusually 
strong 10 $\mu$m emission features; if the gap interpretation holds true, it
would be a region with a large fraction of evolved disks.

Of the disk clearing mechanisms mentioned above, planet formation is the most 
likely explanation for the existence of a gap between inner and outer optically thick
disk regions. Photoevaporation and MRI-induced inner disk draining would not allow
for a remnant inner, optically thick disk region \citep[see also][]{espaillat08}, 
and it is unlikely that dust grain growth would be limited to a certain, well-defined 
radial range in the disk. Since a planet opening a gap will likely migrate inwards on 
roughly the viscous diffusion timescale of $\sim$ 10$^4$ initial orbital periods 
\citep[e.g.,][]{lin86, nelson00}, and the duration of the gap phase as estimated 
above could be at least a factor of a few larger, it is possible that in some systems 
we would be witnessing the effects of already the second or even later generation 
of planets. This type of scenario has also been proposed by \citet{watson09} to
explain the variety of crystalline silicate abundances observed in T Tauri stars in
Taurus. If we assume that at least some of the objects we studied have a disk 
gap opened by a planet, then these planets must have accumulated a mass of
at least a few tens of Earth masses in $\lesssim$ 1 Myr such that they are massive 
enough to open up gaps \citep[see, e.g.,][]{takeuchi96}. Gravitational instability
can create massive planets on a timescale of $\sim$ 10$^3$ years 
\citep[e.g.,][]{boss00}, but also a migrating protoplanet can grow to a
sufficiently large mass within $\sim$ 10$^5$ years by accretion of planetesimals 
\citep{tanaka99}.

A few objects with large $n_{13-31}$ values also have large 10 $\mu$m equivalent
widths (e.g., DM Tau, GM Aur). The large spectral index suggests an inwardly truncated
optically thick disk, with a disk wall at the inner radius. For cases where the presence
of a near-infrared excess indicates that optically thin dust lies in the inner regions,
the unusual strength of the 10 $\mu$m feature can be explained by emission from
this dust \citep[see][]{calvet05}. On the other hand, for objects with no excess 
below about 8 $\mu$m and therefore dust-depleted inner regions, the 10 $\mu$m 
emission likely arises in the atmosphere of the wall \citep{dalessio05,calvet05}. Thus, in
a few cases, not only the object's location in $n_{13-31}$--$EW(10\,{\mu}\mathrm{m})$ 
space, but also the detailed shape of the SED are necessary to interpret disk structure.

\subsection{Median Infrared Excess}

The interpretation of disk evolution based on the median IRS spectrum is complicated
by issues involving normalization. If the distribution of source fluxes in the near-IR
and differences in the distances to the star-forming regions are taken into account, 
then the Class II objects in the Ophiuchus core, Taurus, and Chamaeleon I have 
very similar mid-infrared excess emission; the medians follow each other closely. 
This indicates that, on average, disk structures are similar and that comparable 
amounts of stellar radiation are reprocessed in the disk.

In particular, the Chamaeleon I star-forming region is very similar to Taurus in its 
disk evolution, as gauged from both the median IRS spectra of T Tauri stars and 
the fraction of objects with unusual 13-31 spectral slopes and 10 $\mu$m equivalent 
widths. Agreement between these two regions is expected since their molecular cloud
environments are comparable and their age difference is minimal, especially considering 
that there is an age spread in both regions \cite[e.g.,][]{palla00}, which might just 
be a result of observational uncertainties \citep{hartmann01}.

Based on the distribution of near-IR fluxes, we found that the stars in Chamaeleon I 
are intrinsically fainter, while those in L1688 are brighter than their counterparts in
Taurus. The latter observation can partly be explained by the somewhat higher
fraction of earlier-type stars in our Ophiuchus sample as compared to the other two.
In addition, since the Ophiuchus core region is younger than Taurus, we would 
expect more luminous young stars \citep[e.g.,][]{baraffe02}. The age of 0.3-1 
Myr for the Ophiuchus core was derived assuming a distance of 165 pc 
\citep{luhman99}; the new distance determination of 120 pc by \citet{loinard08} 
implies that stars are a factor of 1.9 less luminous. Assuming they are contracting 
along Hayashi tracks at roughly constant effective temperature ($L \propto 
t^{-2/3}$; \citealt{hartmann98}), this would imply an increase in age by a 
factor of 2.6. Therefore, the median age of $\sim$ 0.3 Myr found for Ophiuchus 
by \citet{luhman99} would change to $\sim$ 0.8 Myr, making it just slightly 
younger than Taurus. The somewhat older age of Chamaeleon I, as compared 
to the other regions, could explain the reduced luminosity of its stars. 

The shape of the median 10 $\mu$m silicate emission feature in the three regions
suggests that the disks in L1688 and Chamaeleon I might have, on average, more 
crystalline silicates than those in Taurus. Since some processing at high temperatures 
is required to transform the amorphous silicate grains from the interstellar medium into 
crystalline form \citep[e.g.,][]{harker02}, we would expect an older star-forming
region, which had more time for this processing to take place, to have a larger
fraction of crystalline grains.
On the other hand, the larger degree of dust processing in the youngest region of 
our sample, L1688, could be a result of different initial conditions in the dense Ophiuchus 
core region. If the mass accretion rates or the masses of the disks were higher, disks 
would likely evolve faster. This could also explain the large fraction of settled disks 
(those with $n_{13-31} \lesssim 0$) in L1688, which is basically the same as in the 
slightly older Taurus region. The fact that our Ophiuchus core sample seems to be 
biased towards somewhat more massive, young stars supports the idea of faster 
disk evolution, given that there are indications that intermediate-mass stars evolve 
faster than low-mass stars \citep{hernandez05}.

\section{Conclusions}
\label{conclusions}

We analyzed the IRS spectra of Class II objects in the Ophiuchus core, Ophiuchus
off-core, Taurus, and Chamaeleon I star-forming regions.
Our main conclusions are as follows:

$\bullet$ The spectral index between 13 and 31 $\mu$m, $n_{13-31}$,
and the equivalent width of the 10 $\mu$m silicate emission feature, 
$EW(10\,{\mu}\mathrm{m})$, are useful indicators for the degree
of disk evolution. We interpret Class II objects with unusually large $n_{13-31}$
as transitional disks with inner disk clearings, while objects with very substantial
$EW(10\,{\mu}\mathrm{m})$ and decreased near- and mid-infrared excess
emission could have gaps in their disks that are filled with optically thin dust 
\citep[see also][]{espaillat07b}. Considering various disk clearing mechanisms, 
the most likely explanation for gaps within optically thick disks is planet formation.

$\bullet$ The Ophiuchus off-core region, which is of similar age as Chamaeleon I 
($\sim$ 2 Myr), has a marginally larger fraction of Class II objects whose 
$n_{13-31}$ and $EW(10\,{\mu}\mathrm{m})$ are outside the range 
found for typical accretion disks, as compared to the other three regions. 
However, the sample size is small, and therefore, if these outliers are interpreted 
as objects with an evolved disk structure, we can only tentatively conclude that 
the Ophiuchus off-core is the most evolved region.

$\bullet$ The transitional disk fraction amounts to a few \% in all four regions;
while L1688 lacks any disks with inner regions fully depleted in small dust grains, 
the frequency of transitional disks in Chamaeleon I is comparable to that of the 
Taurus region. Assuming disk evolution is continuous and affects the structure of 
the inner disk at some point in the lifetime of a disk, we infer that the transitional 
disk phase lasts for several 10$^4$ up to $\sim$ 10$^5$ years.

$\bullet$ The shape and normalized flux levels of the median mid-infrared spectrum 
of disks in Taurus and Chamaeleon I are very similar, and the fractions of objects with 
unusually large $EW(10\,{\mu}\mathrm{m})$, but typical $n_{13-31}$ values, 
are essentially the same. This suggests that the similar star-forming environments and 
age ranges of these two regions result in an overall comparable degree of disk evolution.

$\bullet$ The median mid-infrared excess of T Tauri stars in the Ophiuchus core 
region is also similar to that in Taurus and Chamaeleon I. The shape of the median 
10 $\mu$m silicate emission feature hints at the presence of larger grains and 
crystalline silicates, and therefore dust grains already experienced some processing.
This region also has a sizable fraction ($\sim$ 30\%) of Class II objects with large 
10 $\mu$m equivalent widths, which could indicate that disk evolution sets in early. 

We are likely witnessing the combined effect of age and molecular cloud environment 
on disk evolution. The inner disks in the Taurus, Chamaeleon I, and Ophiuchus off-core 
regions, but also in the younger and denser Ophiuchus core, show clear signs of 
evolution, like the presence of larger grains and decreased flaring, indicative of dust 
sedimentation. Some disks also display evidence for structure formation, like the 
development of inner holes and possibly also gaps. This suggests that grain growth, 
settling, and crystallization occur fast, but the development of radial structure in a
disk requires, on average, more time and might be a transient phenomenon. 
Our analysis implies that the degree of disk evolution does not necessarily correlate 
with age during the first few million years, since it sets in relatively early, and the
subsequent evolution might be dictated more by the individual system properties
than some universal processes happening at set time intervals.

\acknowledgments
We thank the referee for a thoughtful review that led us to improve this paper.
This work is based on observations made with the {\it Spitzer Space Telescope}, 
which is operated by the Jet Propulsion Laboratory (JPL), California Institute of 
Technology (Caltech), under NASA contract 1407. Support for this work was provided 
by NASA through contract number 1257184 issued by JPL/Caltech. 
E.F. was partly supported by a NASA Postdoctoral Program Fellowship, administered 
by Oak Ridge Associated Universities through a contract with NASA, and partly 
supported by NASA through the Spitzer Space Telescope Fellowship Program, through 
a contract issued by JPL/Caltech under a contract with NASA. N. C. and L. H. acknowledge 
support from NASA Origins grants NNG05GI26G, NNG06GJ32G, and NNX08AH94G.
P.D. acknowledges grants from CONACyT, M\'exico.
This publication makes use of data products from the Two Micron All Sky Survey, 
which is a joint project of the University of Massachusetts and the Infrared Processing 
and Analysis Center/Caltech, funded by NASA and the NSF. It has also made use of 
the SIMBAD and VizieR databases, operated at CDS (Strasbourg, France), NASA's 
Astrophysics Data System Abstract Service, and of the NASA/ IPAC Infrared Science 
Archive operated by JPL, Caltech, under contract with NASA. 

Facilities: \facility{Spitzer(IRS)}

\begin{appendix}

Additional tables are provided in the electronic edition. To show their format and 
content type, the first few lines of these tables are printed here. Table 5 contains
the target names and coordinates of our sample of Class II objects in Taurus,
Chamaeleon I, and the Ophiuchus core and off-core regions. Table 6 contains
the median IRS spectra and their lower and upper quartiles for the Ophiuchus 
core, Chamaeleon I, and Taurus regions, normalized at the 2MASS H-band fluxes. 
Here, only the first few entries of the Ophiuchus core median and quartiles are shown. 
These are the medians displayed in Figure \ref{Tau_Cha_Oph_median} as solid lines. 

\begin{deluxetable}{lcc}
\tablecaption{Target names and coordinates}
\tablenum{5}
\tablehead{\colhead{Name} & \colhead{R.A. (J2000)} & \colhead{Dec. (J2000)}
}
\startdata
04108+2910 & 04 13 57.38 & 29 18 19.3 \\
04187+1927 & 04 21 43.24 & 19 34 13.3 \\
04200+2759 & 04 23 07.77 & 28 05 57.3 \\
04216+2603 & 04 24 44.58 & 26 10 14.1 \\
04303+2240 & 04 33 19.07 & 22 46 34.2 
\enddata
\tablecomments{Table 5 is published in its entirety in the electronic 
edition of the {\it Astrophysical Journal}.  A portion is shown here for 
guidance regarding its form and content.}
\end{deluxetable}

\begin{deluxetable}{lccc}
\tablecaption{Ophiuchus core median (H normalization)}
\tablenum{6}
\tablehead{\colhead{$\lambda$} & \colhead{$\nu$F$_{\nu}$} & 
\colhead{$\nu$F$_{\nu,lower}$} & \colhead{$\nu$F$_{\nu,upper}$} \\ 
\colhead{($\mu$m)} & \colhead{(erg/s/cm$^2$)} & \colhead{(erg/s/cm$^2$)} & 
\colhead{(erg/s/cm$^2$)} } 
\startdata
5.0 & 4.960E-10 & 3.300E-10 & 7.499E-10 \\
5.1 & 4.561E-10 & 2.778E-10 & 7.436E-10 \\
5.2 & 4.002E-10 & 2.350E-10 & 5.905E-10 \\
5.3 & 3.665E-10 & 2.123E-10 & 5.743E-10 \\
5.4 & 3.509E-10 & 2.043E-10 & 5.385E-10 \\
5.5 & 3.241E-10 & 1.907E-10 & 4.998E-10 
\enddata
\tablecomments{Table 6 is published in its entirety in the electronic 
edition of the {\it Astrophysical Journal}.  A portion is shown here for 
guidance regarding its form and content.}
\end{deluxetable}

\end{appendix}

\clearpage

\LongTables
\begin{deluxetable}{lccccc}
\tabletypesize{\small}  
\tablecaption{Properties of the targets in the Ophiuchus core region
\label{disk_evol_Oph-core}}
\tablehead{
\colhead{Name} & \colhead{Spectral Type$^a$} & \colhead{$A_V$$^b$} & 
\colhead{$EW(10\,{\mu}\mathrm{m})$} & \colhead{$n_{13-31}$} & 
\colhead {Median} \\
\colhead{(1)} & \colhead{(2)} & \colhead{(3)} & \colhead{(4)} & \colhead{(5)} &
\colhead{(6)} 
}
\startdata        
     16220-2452 &    M3 &   3.7 &  4.87 & -0.09 &   \\        
     16237-2349 &  K5.5 &   5.0 &  1.67 & -0.17 &  $\times$ \\
 B162713-241818 &    M0 &  13.0 &  1.39 & -1.28 &  $\times$ \\
         DoAr24 &    K5 &   4.1 &  3.78 & -0.39 &  $\times$ \\
          GSS29 &    M1 &  11.5 &  2.65 & -1.15 &  $\times$ \\
          GSS31 &    G6 &   7.0 &  1.58 & -0.50 &   \\        
          GSS37 &    M0 &   8.7 &  1.47 & -1.23 &  $\times$ \\
          GSS39 &    M0 &  17.2 &  1.15 & -0.75 &  $\times$ \\
          GY144 &    M5 &  40.3 &  3.23 & -0.86 &   \\        
          GY154 &    M6 &  23.3 &  2.63 & -0.27 &   \\        
          GY188 &    M3 &  40.7 &  1.85 &  0.15 &   \\        
          GY195 &    M3 &  24.7 &  6.23 & -0.15 &   \\        
          GY204 &  M5.5 &   3.4 &  3.41 &  0.22 &   \\        
          GY213 &    M4 &  27.3 &  3.21 & -0.95 &   \\        
          GY224 &    M4 &  34.2 &  3.94 & -0.49 &   \\        
          GY235 &    M5 &   8.7 &  3.02 &  0.14 &   \\        
          GY245 &   \nodata &  44.2 &  3.53 & -0.72 &   \\        
          GY260 &    M4 &  31.7 &  2.31 & -0.18 &   \\      
          GY284 & M3.25 &   7.0 &  3.60 &  0.08 &   \\        
          GY289 &    M2 &  24.0 &  1.54 & -0.20 &  $\times$ \\
          GY292 &    K7 &  13.5 &  6.31 & -1.41 &  $\times$ \\
          GY301 &    K7 &  40.4 &  3.15 & -0.26 &  $\times$ \\
          GY310 &    M4 &   7.5 &  2.92 &  0.31 &   \\        
          GY314 &    K5 &   8.1 &  1.08 & -0.20 &  $\times$ \\
          GY323 &    M5 &  32.8 &  2.99 & -1.17 &   \\        
          GY326 &    M2 &  10.2 &  2.44 & -0.48 &  $\times$ \\
         GY344 &    M6 &  19.1 &  1.35 &  0.02 &   \\        
          GY350 &    M6 &   8.5 &  2.53 & -0.94 &   \\        
          GY352 &    M5 &  20.3 &  0.91 & -1.10 &   \\        
          GY371 &    M6 &   6.3 &  0.44 & -0.71 &   \\        
          GY397 &    M6 &   5.8 &  1.18 & -0.76 &   \\        
       HD 147889 &    B2 &   4.4 &  0.02 & -0.11 &   \\        
           IRS2 &  K3.5 &   7.7 &  1.53 & -0.23 &   \\        
          IRS26 &    M6 &  19.6 &  3.41 & -0.78 &   \\        
           IRS3 &    M2 &   8.5 &  6.56 &  0.37 &  $\times$ \\
          IRS33 &    M2 &  40.5 &  0.99 & -1.08 &  $\times$ \\
          IRS34 &    M0 &  28.7 &  0.85 & -0.94 &  $\times$ \\
          IRS35 &   \nodata &  45.7 &  3.65 & -1.03 &   \\      
          IRS42 &    K7 &  29.0 &  3.08 & -0.90 &  $\times$ \\
          IRS45 &  K6.5 &  24.5 &  1.59 & -0.87 &  $\times$ \\
          IRS47 &    M3 &  28.0 &  1.66 & -1.33 &   \\        
          IRS48 &    A0 &  15.0 & -0.30 &  1.33 &   \\        
          IRS49 &    K5 &  12.2 &  3.67 & -0.11 &  $\times$ \\
          IRS51 &    K6 &  37.6 &  5.53 & -0.64 &  $\times$ \\
            S2 &    M0 &  11.8 &  0.55 & -1.33 &  $\times$ \\
            SR4 &  K4.5 &   2.6 &  2.29 &  0.29 &   \\        
            SR9 &    K5 &   1.9 &  5.87 &  0.02 &  $\times$ \\
           SR10 &    M2 &   1.5 &  3.03 & -1.34 &  $\times$ \\
           SR20 &    G7 &   7.0 &  1.21 & -2.53 &   \\        
          SR21 &    F4 &   8.0 &  4.01 &  1.57 &   \\        
          VSS27 &  G3.5 &   8.1 &  1.53 & -0.97 &   \\        
          VSSG1 &    M0 &  16.5 &  0.88 & -1.25 &  $\times$ \\
          VSSG5 &    M0 &  20.3 &  1.00 & -1.46 &  $\times$ \\
        VSSG25 &    M4 &  13.6 &  3.38 & -0.21 &   \\        
           WL1 &    M4 &  28.2 &  4.51 & -0.19 &   \\        
          WL2 &    M0 &  34.8 &  2.63 & -0.68 &  $\times$ \\
            WL3 &    M4 &  42.0 &  1.34 & -0.42 &   \\        
            WL4 &  M1.5 &  21.3 &  1.77 &  0.06 &  $\times$ \\
           WL10 &    K7 &  13.3 &  1.08 & -0.31 &  $\times$ \\
           WL11 &    M0 &  15.9 &  1.97 & -0.41 &  $\times$ \\
           WL18 &  K6.5 &  12.2 &  1.40 & -0.34 &  $\times$ \\
          WSB37 &    M5 &   2.9 &  1.23 & -0.88 &   \\        
          WSB60 &  M4.5 &   4.5 &  2.01 &  0.05 &   
\enddata
\tablecomments{Column (1) lists the name of the object, column (2) the
adopted spectral type, column (3) the optical extinction $A_V$ (assuming $R_V$=3.1), 
column (4) the 10 $\mu$m equivalent width ($EW(10\,{\mu}\mathrm{m})$), 
column (5) the 13-31 $\mu$m spectral index ($n_{13-31}$), and the $\times$ 
symbol in column (6) indicates that the object was used in the median calculation. 
Note that uncertainties in $A_V$, ranging from 1\% to 30\% of the $A_V$ value
(McClure et al., in preparation), account for $\sim$ 25\% of the error bars of 
$EW(10\,{\mu}\mathrm{m})$ and $n_{13-31}$. \\
$^a$ Spectral types were taken from \citet{brandner97,luhman99,wilking05,natta06}; 
McClure et al. (in preparation). \\
$^b$ $A_V$ values are from McClure et al. (in preparation).}
\end{deluxetable}

\LongTables
\begin{deluxetable}{lccccc}
\tabletypesize{\small}  
\tablecaption{Properties of the targets in the Taurus region 
\label{disk_evol_Tau}}
\tablehead{
\colhead{Name} & \colhead{Spectral Type$^a$} & \colhead{$A_V$$^a$} &
\colhead{$EW(10\,{\mu}\mathrm{m})$} & \colhead{$n_{13-31}$} & \colhead {Median} \\
\colhead{(1)} & \colhead{(2)} & \colhead{(3)} & \colhead{(4)} & \colhead{(5)} &
\colhead{(6)} 
}
\startdata
     04108+2910 &    M0 &   1.4 &  0.28 & -0.79 &  $\times$ \\
     04187+1927 &    M0 &   0.0 &  0.94 & -0.84 &  $\times$ \\
     04200+2759 &   \nodata &   0.0 &  1.52 & -0.41 &   \\        
     04216+2603 &    M1 &   0.0 &  1.16 & -0.41 &  $\times$ \\
     04303+2240 &   \nodata &  11.7 &  2.21 & -1.15 &   \\        
     04370+2559 &   \nodata &   9.8 &  4.06 & -0.51 &   \\        
     04385+2550 &    M0 &   7.8 &  2.48 &  0.26 &  $\times$ \\
          AA Tau &    K7 &   1.8 &  1.79 & -0.51 &  $\times$ \\
          AB Aur &    A0 &   0.3 &  4.07 &  0.81 &   \\        
          BP Tau &    K7 &   1.0 &  2.64 & -0.58 &  $\times$ \\
          CI Tau &    K7 &   2.0 &  2.55 & -0.17 &  $\times$ \\
       CoKu Tau/3 &    M1 &   5.0 &  2.66 & -1.19 &  $\times$ \\
       CoKu Tau/4 &    M1.5 &   3.0 &  5.37 &  2.12 &  $\times$ \\
          CW Tau &    K3 &   2.8 &  1.15 & -0.65 &   \\        
          CX Tau &    M0 &   1.3 &  2.30 & -0.15 &  $\times$ \\
          CY Tau &    K7 &   1.7 &  0.77 & -0.99 &  $\times$ \\
          CZ Tau &    M1.5 &   2.4 &  2.40 & -1.04 &  $\times$ \\
          DD Tau &    M3.5 &   1.0 &  1.10 & -0.74 &   \\        
         DE Tau &    M0 &   1.2 &  1.81 & -0.13 &  $\times$ \\
          DF Tau &    M0 &   1.6 &  0.83 & -1.09 &  $\times$ \\
          DG Tau &   \nodata  &   1.6 &  0.13 &  0.15 &   \\        
          DH Tau &    M0 &   1.7 &  3.57 &  0.26 &  $\times$ \\
          DK Tau &    M0 &   1.3 &  3.79 & -0.80 &  $\times$ \\
          DL Tau &    K7 &   1.5 &  0.51 & -0.77 &  $\times$ \\
       DM Tau &    M1 &   0.7 &  5.80 &  1.29 &  $\times$ \\
          DN Tau &    M0 &   0.6 &  1.06 & -0.42 &  $\times$ \\
          DO Tau &    M0 &   2.0 &  0.94 & -0.13 &  $\times$ \\
          DP Tau &    M0.5 &   0.6 &  1.70 & -0.33 &  $\times$ \\
          DQ Tau &    M0 &   1.6 &  0.65 & -0.41 &  $\times$ \\
          DR Tau &   \nodata  &   1.2 &  1.23 & -0.40 &   \\        
          DS Tau &    K5 &   1.1 &  2.35 & -0.96 &  $\times$ \\
    F04101+3103 &    A1 &   1.9 &  4.46 & -0.08 &   \\        
    F04147+2822 &    M4 &   2.5 &  2.55 & -1.11 &   \\        
    F04192+2647 &   \nodata &   0.0 &  1.17 & -0.47 &   \\        
    F04262+2654 &   \nodata &   0.0 &  0.78 & -0.13 &   \\        
   F04297+2246A &   \nodata &   0.0 &  3.93 & -0.41 &   \\        
    F04570+2520 &   \nodata &   0.0 &  0.94 & -1.60 &   \\        
          FM Tau &    M0 &   1.4 &  3.13 & -0.16 &  $\times$ \\
          FN Tau &    M5 &   1.4 &  1.84 & -0.05 &   \\        
          FO Tau &    M2 &   3.0 &  1.40 & -0.32 &  $\times$ \\
          FP Tau &    M4 &   0.0 &  0.94 & -0.10 &   \\        
          FQ Tau &    M2 &   1.9 &  1.23 & -0.46 &  $\times$ \\
          FS Tau &    M1 &   1.4 &  1.28 &  0.08 &  $\times$ \\
          FT Tau &     C &   0.0 &  1.86 & -0.46 &   \\        
          FV Tau &    K5 &   5.3 &  1.47 & -0.66 &  $\times$ \\
          FX Tau &    M1 &   2.0 &  4.04 & -0.39 &  $\times$ \\
          FZ Tau &    M0 &   3.7 &  1.26 & -0.87 &  $\times$ \\
          GG Tau &    M0 &   1.0 &  2.82 & -0.31 &  $\times$ \\
          GH Tau &    M2 &   1.0 &  1.42 & -0.33 &  $\times$ \\
          GI Tau &    K6 &   2.3 &  2.65 & -0.63 &  $\times$ \\
          GK Tau &    M0 &   1.1 &  4.60 & -0.37 &  $\times$ \\
          GM Aur &    K3 &   1.2 &  5.07 &  1.75 &   \\        
          GN Tau &    M2 &   3.5 &  2.93 & -1.01 &  $\times$ \\
          GO Tau &    M0 &   2.0 &  1.99 &  0.03 &  $\times$ \\
       Haro 6-13 &    M0 &  11.9 &  3.87 &  0.38 &  $\times$ \\
       Haro 6-37 &    K7 &   3.8 &  1.38 & -0.85 &  $\times$ \\
          HK Tau &    M0.5 &   2.7 &  1.80 &  0.74 &  $\times$ \\
          HN Tau &    K5 &   1.5 &  2.44 & -0.45 &  $\times$ \\
          HO Tau &    M0.5 &   1.3 &  2.46 & -0.63 &  $\times$ \\
          HP Tau &    K3 &   2.8 &  2.43 & -0.01 &   \\        
          HQ Tau &   \nodata &   0.0 &  3.05 & -0.64 &   \\        
          IP Tau &    M0 &   0.5 &  4.18 & -0.09 &  $\times$ \\
          IQ Tau &    M0.5 &   1.4 &  1.75 & -1.00 &  $\times$ \\
          IS Tau &    M0 &   3.2 &  2.30 & -1.21 &  $\times$ \\
          IT Tau &    K2 &   3.8 &  1.12 & -0.87 &   \\        
         LkCa 15 &    K5 &   1.2 &  6.70 &  0.62 &  $\times$ \\
           MHO 3 &    K7 &   8.3 &  5.93 &  0.27 &  $\times$ \\
         RW Aur A &    K3 &   0.5 &  1.25 & -0.54 &   \\        
          RY Tau &    G1 &   2.2 &  6.51 & -0.09 &   \\        
         SU Aur &    G1 &   0.9 &  4.75 &  0.74 &   \\        
           T Tau &    K0 &   1.8 & -0.27 &  0.55 &   \\        
         UX Tau A &    K5 &   0.7 &  0.78 &  1.83 &  $\times$ \\
          UY Aur &    K7 &   2.1 &  2.05 & -0.04 &  $\times$ \\
        UZ Tau &    M1 &   1.0 &  2.25 & -0.73 &  $\times$ \\
     V410 Anon 13 &    M6 &   5.8 &  1.70 & -0.58 &   \\        
        V710 Tau &    M1 &   1.9 &  1.43 & -0.70 &  $\times$ \\
     V773 Tau &    K3 &   2.0 &  0.82 & -0.85 &   \\        
       V807 Tau &    K7 &   0.6 &  0.67 & -0.18 &  $\times$ \\
        V836 Tau &    K7 &   1.1 &  3.34 & -0.45 &  $\times$ \\
        V892 Tau &    B9 &   8.0 &  3.06 & -0.13 &   \\        
        V955 Tau &    K5 &   3.7 &  1.54 & -0.94 &  $\times$ \\
          VY Tau &    M0 &   1.4 &  2.34 & -0.13 &  $\times$ \\
          XZ Tau &    M2 &   2.9 &  0.66 & -0.32 &  $\times$ \\
     ZZ Tau &    M3 &   1.4 &  1.01 & -0.89 &   \\
       ZZ Tau IRS &    M4.5 &   1.5 &  1.58 &  0.17 &   
\enddata
\tablecomments{See the footnote of Table \ref{disk_evol_Oph-core} for an
explanation of the table content. 
Uncertainties in $A_V$, assumed to be 0.3 mag \citep{kenyon95}, account for 
$\sim$ 15\% and 1\% of the error bars of $EW(10\,{\mu}\mathrm{m})$ 
and $n_{13-31}$, respectively.\\
$^a$ Spectral type and extinction information were taken from \citet{kenyon90,
hartigan94,strom94,kenyon95,torres95,kenyon98,briceno98,luhman00,white01,
hartigan03,white03,white04,calvet04,furlan06}.}
\end{deluxetable}

\LongTables
\begin{deluxetable}{lccccc}
\tabletypesize{\small}  
\tablecaption{Properties of the targets in the Chamaeleon I region 
\label{disk_evol_Cha}}
\tablehead{
\colhead{Name} & \colhead{Spectral Type$^a$} & \colhead{$A_V$$^a$} &
\colhead{$EW(10\,{\mu}\mathrm{m})$} & \colhead{$n_{13-31}$} & \colhead {Median} \\
\colhead{(1)} & \colhead{(2)} & \colhead{(3)} & \colhead{(4)} & \colhead{(5)} &
\colhead{(6)} 
}
\startdata
2M J10580597-7711501 & M5.25 &   1.6 &  3.22 &  0.23 &   \\ 
2M J11062942-7724586 &   M6 &  20.0 &  1.53 & -0.27 &   \\           
2M J11065939-7530559 & M5.25 &   0.6 &  1.70 &  0.37 &   \\    
2M J11070369-7724307 &  M7.5 &  16.5 &  2.60 &  0.19 &   \\
2M J11241186-7630425 &    M5 &   2.7 &  4.52 &  0.68 &   \\
    B43 & M3.25 &   8.0 &  6.98 &  0.42 &   \\  
    C1-6 & M1.25 &  11.6 &  2.34 & -0.55 &  $\times$ \\
    C7-1 &    M5 &   5.0 &  1.08 & -0.54 &   \\       
  Cha Ha1 & M7.75 &   0.0 &  1.87 &  0.00 &   \\        
  Cha Ha2 & M5.25 &   3.8 &  1.08 & -1.00 &   \\                   
   CHSM 7869 &    M6 &   1.6 &  1.74 & -0.57 &   \\                       
   CHSM 10862 & M5.75 &   1.6 &  0.17 & -0.06 &   \\ 
   CHXR 20 &    K6 &   3.5 &  4.28 & -0.99 &  $\times$ \\
   CHXR 22E &  M3.5 &   1.4 &  0.22 &  0.63 &   \\ 
   CHXR 30A &    M0 &  10.6 &  2.66 & -0.60 &  $\times$ \\
   CHXR 30B & M1.25 &  11.2 &  2.99 & -0.40 &  $\times$ \\
   CHXR 47 &    K3 &   5.1 &  2.23 & -0.66 &   \\       
   CR Cha &    K2 &   1.5 &  4.66 & -0.14 &   \\     
   CS Cha &    K6 &   0.3 &  3.02 &  2.89 &  $\times$ \\
   CT Cha &    K5 &   1.6 &  2.14 & -0.31 &  $\times$ \\             
   CU Cha &  B9.5 &   1.5 & -0.49 &  1.04 &   \\        
   CV Cha &    G9 &   1.5 &  5.78 & -0.27 &   \\   
   DI Cha &    G2 &   2.7 &  1.91 & -0.67 &   \\ 
   Hn 5 &  M4.5 &   1.1 &  2.07 & -1.45 &   \\   
   Hn 10E & M3.25 &   3.6 &  3.23 &  0.35 &   \\                    
   Hn 11 &    M0 &   7.6 &  2.56 & -0.73 &  $\times$ \\
   Hn 13 & M5.75 &   0.8 &  1.29 & -0.44 &   \\        
   Hn 21W &    M4 &   2.6 &  1.45 & -0.77 &   \\    
   ISO 52 &    M4 &   1.3 &  2.20 &  0.11 &   \\
   ISO 79 & M5.25 &   7.9 &  1.58 & -0.14 &   \\  
   ISO 91 &    M3 &  14.5 &  7.53 & -0.40 &   \\                      
   ISO 138 &    M7 &   0.0 &  1.81 & -0.27 &   \\        
   ISO 143 &    M5 &   3.0 &  1.47 & -1.01 &   \\        
   ISO 220 & M5.75 &   6.1 &  1.44 & -0.82 &   \\          
   ISO 225 & M1.75 &   4.4 &  1.11 &  0.15 &  $\times$ \\
   ISO 235 &  M5.5 &   7.5 &  1.48 & -1.26 &   \\      
   ISO 237 &  K5.5 &   6.8 &  3.43 &  0.51 &  $\times$ \\
   ISO 252 &    M6 &   3.4 &  1.35 & -0.85 &   \\ 
   ISO 256 &  M4.5 &   9.1 &  3.27 & -0.63 &   \\    
   ISO 282 & M4.75 &   3.6 &  1.86 & -0.53 &   \\        
  SX Cha &    M0 &   2.8 &  2.52 & -0.48 &  $\times$ \\      
  SZ Cha &    K0 &   1.5 &  3.38 &  1.68 &   \\    
  T5 & M3.25 &   1.2 &  1.36 & -0.32 &   \\      
  T21 &    G5 &   3.3 & -0.04 & -1.05 &   \\ 
  T25 &  M2.5 &   1.6 &  3.30 &  2.79 &  $\times$ \\
  T28 &    M0 &   4.8 &  2.05 & -0.60 &  $\times$ \\
  T29 &    K6 &   7.3 &  2.03 & -0.34 &  $\times$ \\
  T33A &    G7 &   3.0 &  5.38 & -0.54 &   \\  
  T35 &    M0 &   3.5 &  0.51 &  1.49 &  $\times$ \\
  T42 &    K5 &   8.2 &  1.46 & -0.02 &  $\times$ \\
  T43 &    M2 &   5.2 &  1.90 & -0.41 &  $\times$ \\
  T45a &    M1 &   1.9 &  1.66 &  0.10 &  $\times$ \\
  T47 &    M2 &   4.2 &  2.14 & -0.28 &  $\times$ \\
  T50 &    M5 &   0.4 &  2.04 & -0.32 &   \\        
  T51 &  K3.5 &   0.8 &  4.18 & -1.53 &   \\        
  T54 &    G8 &   1.8 &  0.62 &  1.10 &   \\               
  T56 &  M0.5 &   0.6 &  7.57 &  0.90 &  $\times$ \\
  TW Cha &    M0 &   1.2 &  5.57 & -0.18 &  $\times$ \\
  UY Cha & M4.25 &   0.0 &  3.10 & -0.93 &   \\        
  UZ Cha &  M0.5 &   2.2 &  2.13 & -0.22 &  $\times$ \\
  VV Cha &    M3 &   0.5 &  2.43 & -0.79 &   \\              
  VW Cha &    M0 &   2.6 &  2.52 & -0.21 &  $\times$ \\
  VY Cha &  M0.5 &   3.2 &  2.58 & -0.22 &  $\times$ \\
  VZ Cha &    K6 &   2.0 &  1.35 & -0.91 &  $\times$ \\
  WW Cha &    K5 &   4.8 &  4.16 &  0.42 &  $\times$ \\
  WX Cha & M1.25 &   2.0 &  2.56 & -1.03 &  $\times$ \\
  WY Cha &    M0 &   3.0 &  1.61 & -1.18 &  $\times$ \\
  WZ Cha & M3.75 &   0.5 &  1.97 & -0.75 &   \\        
  XX Cha &    M2 &   1.2 &  1.77 & -0.39 &  $\times$ 
\enddata
\tablecomments{See the footnote of Table \ref{disk_evol_Oph-core} for an
explanation of the table content. 
Uncertainties in $A_V$, assumed to be 0.46 mag \citep{luhman07}, account for 
$\sim$ 10\% and 2\% of the error bars of $EW(10\,{\mu}\mathrm{m})$ 
and $n_{13-31}$, respectively. \\
$^a$ Spectral type and, where available, extinction information were taken from 
\citet{luhman04a} and \citet{gomez03}. For objects without published or 
with uncertain extinction values, the optical extinction $A_V$ was derived using 
the observed optical/near-infrared colors and intrinsic colors based on the spectral 
type or a typical CTTS excess (see text for details).}
\end{deluxetable}

\LongTables
\begin{deluxetable}{lcccc}
\tabletypesize{\small}  
\tablecaption{Properties of the targets in the Ophiuchus off-core region 
\label{disk_evol_Oph-off}}
\tablehead{
\colhead{Name} & \colhead{Spectral Type$^a$} & \colhead{$A_V$$^b$} &
\colhead{$EW(10\,{\mu}\mathrm{m})$} & \colhead{$n_{13-31}$} \\
\colhead{(1)} & \colhead{(2)} & \colhead{(3)} & \colhead{(4)} & \colhead{(5)}
}
\startdata
     16126-2235 &    M3 &   0.6 & 10.98 &  1.15 \\
     16156-2358 &    F0 &   2.2 & -0.02 &  1.36 \\
     16193-2314 &    G5 &   3.5 &  3.47 & -0.14 \\
     16201-2410 &    G0 &   8.0 &  6.40 &  1.32 \\
     16225-2607 &    K7 &   1.5 &  6.89 &  0.52 \\
     16289-2457 &    G5 &   9.5 &  5.16 &  0.21 \\
     16293-2424 &    G0 &  17.1 &  2.49 & -0.03 \\
       DoAr 16 &    K6 &   3.8 &  2.30 & -0.03 \\
       DoAr 28 &    K5 &   2.7 &  1.85 &  2.19 \\
         IRS 60 &    K2 &   9.0 &  3.24 & -0.49 \\
      L 1689SNO2 &    M3 &  15.0 &  0.45 & -0.83 \\
        ROX 42C &    K6 &   1.9 &  2.08 & -0.37 \\
       ROX 43A1 &    G0 &   3.5 &  5.86 & -0.52 \\
         ROX 44 &    K0 &   3.8 &  8.05 &  0.63 \\
        ROX 47A &    M3 &   1.9 &  1.20 & -0.66 
\enddata
\tablecomments{See the footnote of Table \ref{disk_evol_Oph-core} for an
explanation of the table content. 
Uncertainties in $A_V$, which are typically 0.3 mag (McClure et al., in preparation), 
account for $\sim$ 10\% and 1\% of the error bars of 
$EW(10\,{\mu}\mathrm{m})$ and $n_{13-31}$, respectively. \\
$^a$ Spectral types were taken from \citet{herbig88,chen95,preibisch98,vieira03,
torres06}; McClure et al. (in preparation). \\
$^b$ $A_V$ values are from McClure et al. (in preparation).}
\end{deluxetable}


\clearpage

\begin{thebibliography}{}
\bibitem[Adams et al.(1987)]{adams87} Adams, F. C., Lada, Ch. J., \& Shu, F. H.  
1987, \apj, 312, 788
\bibitem[Alexander et al.(2006)]{alexander06} Alexander, R. D., Clarke, C. J.,
\& Pringle, J. E.  2006, \mnras, 369, 229
\bibitem[Alexander \& Armitage(2007)]{alexander07} Alexander, R. D., \& 
Armitage, P. J.  2007, \mnras, 375, 500
\bibitem[Andr{\'e} \& Montmerle(1994)]{andre94} Andr{\'e}, P., \& Montmerle, 
T. 1994, \apj, 420, 837
\bibitem[Andrews et al.(2009)]{andrews09} Andrews, S. M., Wilner, D. J., 
Hughes, A. M., Qi, C., \& Dullemond, C. P.  2009, \apj, in press, arXiv:0906.0730
[astro-ph]
\bibitem[Apai et al.(2005)]{apai05} Apai, D., Pascucci, I., Bouwman, J., Natta, A.,
Henning, Th., \& Dullemond, C.  2005, Science, 310, 834
\bibitem[Artymowicz \& Lubow(1994)]{artymowicz94} Artymowicz, P., \&  Lubow, 
S. H.  1994, \apj, 421, 651
\bibitem[Baraffe et al.(2002)]{baraffe02} Baraffe, I., Chabrier, G., Allard, F., \&
Hauschildt, P. H.  2002, \aap, 382, 563
\bibitem[Barsony et al.(1997)]{barsony97} Barsony, M., Kenyon, S. J., Lada, E. A.,
\& Teuben, P. J.  1997, \apjs, 112, 109
\bibitem[Bertout et al.(1999)]{bertout99} Bertout, C., Robichon, N., \& Arenou, F. 
1999, \aap, 352, 574
\bibitem[Bontemps et al.(2001)]{bontemps01} Bontemps, S., Andr\'e, P., Kalas, A. A.,
Nordh, L., Olofsson, G., Huldtgren, M., Abergel, A., Blommaert, J., et al.  2001, \aap,
372, 173
\bibitem[Boss(2000)]{boss00} Boss, A. P.  2000, \apj, 536, L101
\bibitem[Bouwman et al.(2001)]{bouwman01} Bouwman, J., Meeus, G., de Koter, A.,
Hony, S., Dominik, C., \& Waters, L. B. F. M.  2001, \aap, 375, 950
\bibitem[Brandner \& Zinnecker(1997)]{brandner97} Brandner, W., \& Zinnecker, H.
1997, \aap, 321, 220
\bibitem[Brice\~no et al.(1998)]{briceno98} Brice\~no, C., Hartmann, L., Stauffer, J.,
\& Mart{\'\i}n, E.  1998, \aj, 115, 2074
\bibitem[Brice\~no et al.(2002)]{briceno02} Brice\~no, C., Luhman, K. L., Hartmann, L.,
Stauffer, J. R., \& Kirkpatrick, J. D.  2002, \apj, 580, 317
\bibitem[Brown et al.(2007)]{brown07} Brown, J. M., Blake, G. A., Dullemond, C. P.,
Mer{\'\i}n, B., Augereau, J. C., Boogert, A. C. A., Evans, N. J., Geers, V. C., et al.
2007, \apjl, 664, L107
\bibitem[Calvet et al.(2004)]{calvet04} Calvet, N., Muzerolle, J., Brice\~no, C., 
Hern{\'a}ndez, J., Hartmann, L., Saucedo, J. L., \& Gordon, K. D.  2004, \aj, 
128, 1294
\bibitem[Calvet et al.(2005)]{calvet05} Calvet, N., D'Alessio, P., Watson, D. M., 
Franco-Hern{\'a}ndez, R., Furlan, E., Green, J., Sutter, P. M., Forrest, W. J., et al.
2005, \apjl, 630, L185
\bibitem[Cambr{\'e}sy et al.(1998)]{cambresy98} Cambr{\'e}sy, L., Copet, E.,
Epchtein, N., de Batz, B., Borsenberger, J., Fouqu{\'e}, P., Kimeswenger, S., \&
Tiph{\'ene}, D.  1998, \aap, 338, 977
\bibitem[Cardelli et al.(1989)]{cardelli89} Cardelli, J. A., Clayton, G. C., \& Mathis, 
J. S.  1989, \apj, 345, 245
\bibitem[Chen et al.(1995)]{chen95} Chen, H., Myers, P. C., Ladd, E. F., \& Wood,
D. O. S. 1995, \apj, 445, 377
\bibitem[Chen et al.(1997)]{chen97} Chen, H., Grenfell, T. G., Myers, P. C., \&
Hughes, J. D.  1997, \apj, 478, 295
\bibitem[Chiang \& Goldreich(1997)]{chiang97} Chiang, E. I., \& Goldreich, P. 1997,
\apj, 490, 368
\bibitem[Chiang \& Murray-Clay(2007)]{chiang07} Chiang, E., \& Murray-Clay, R.
2007, Nature Physics, 3, 604
\bibitem[Chiar et al.(2007)]{chiar07} Chiar, J. E., Ennico, K., Pendleton, Y. J., Boogert,
A. C. A., Greene, T., Knez, C., Lada, C., Roellig, et al. 2007, \apjl, 666, L73
\bibitem[Clarke et al.(2001)]{clarke01} Clarke, C. J., Gendrin, A., \& Sotomayor, M.  
2001, \mnras, 328, 485
\bibitem[Cohen \& Kuhi(1979)]{cohen79} Cohen, M., \& Kuhi, L. V. 1979, \apjs, 
41, 743
\bibitem[Correia et al.(2006)]{correia06} Correia, S., Zinnecker, H., Ratzka, Th., \&
Sterzik, M. F. 2006, \aap, 459, 909
\bibitem[Currie \& Kenyon(2009)]{currie09} Currie, T., \& Kenyon, S. J.  2009,
\aj, 138, 703
\bibitem[D'Alessio et al.(1999)]{dalessio99} D'Alessio, P., Calvet, N., Hartmann, L., 
Lizano, S., \& Cant\'o, J. 1999, \apj, 527, 893
\bibitem[D'Alessio et al.(2005)]{dalessio05} D'Alessio, P., Hartmann, L., Calvet, N.,
Franco-Hern\'andez, R., Forrest, W. J., Sargent, B., Furlan, E., Uchida, K., et al. 2005, 
\apj, 621, 461
\bibitem[D'Alessio et al.(2006)]{dalessio06} D'Alessio, P., Calvet, N., Hartmann, L., 
Franco-Hern{\'a}ndez, R., \& Serv{\'\i}n, H.  2006, \apj, 638, 314
\bibitem[Dorschner et al.(1995)]{dorschner95} Dorschner, J., Begemann, B.,
Henning, Th., J{\"a}ger, C., \& Mutschke, H. 1995, \aap, 300, 503
\bibitem[Doucet et al.(2007)]{doucet07} Doucet, C., Habart, E., Pantin, E., 
Dullemond, C., Lagage, P. O., Pinte, C., Duch\^ene, G., \& M\'enard, F.  2007,
\aap, 470, 625
\bibitem[Draine \& Lee(1984)]{draine84} Draine, B. T., \& Lee, H. M. 1984,
\apj, 285, 89
\bibitem[Duch\^ene et al.(2003)]{duchene03} Duch\^ene, G., Ghez, A. M.,
McCabe, C., \& Weinberger, A. J.  2003, \apj, 592, 288
\bibitem[Dullemond \& Dominik(2004)]{dullemond04} Dullemond, C. P., \& Dominik, C.
2004, \aap, 421, 1075
\bibitem[Dullemond \& Dominik(2005)]{dullemond05} Dullemond, C. P., \& Dominik, C.
2005, \aap, 434, 971 
\bibitem[Espaillat et al.(2007a)]{espaillat07a} Espaillat, C., Calvet, N., D'Alessio, P.,
Bergin, E., Hartmann, L., Watson, D. M., Furlan, E., Najita, N., et al. 2007a, 
\apjl, 664, L111
\bibitem[Espaillat et al.(2007b)]{espaillat07b} Espaillat, C., Calvet, N., D'Alessio, P.,
Hern{\'a}ndez, J., Qi, C., Hartmann, L., Furlan, E., \& Watson, D. M. 2007b, \apjl,
670, L135
\bibitem[Espaillat et al.(2008)]{espaillat08} Espaillat, C., Calvet, N., Luhman, K. L.,
Muzerolle, J., \& D'Alessio, P.  2008, \apjl, 682, L125
\bibitem[Espaillat(2009)]{espaillat09} Espaillat, C. 2009, Ph.D. thesis, 
University of Michigan  
\bibitem[Fabian et al.(2001)]{fabian01} Fabian, D., Henning, Th., J{\"ager}, C.,
Mutschke, H., Dorschner, J., \& Wehrhan, O. 2001, \aap, 378, 228
\bibitem[Feigelson \& Kriss(1989)]{feigelson89} Feigelson, E. D., \& Kriss, G. A.
1989, \apj, 338, 262
\bibitem[Forrrest et al.(2004)]{forrest04} Forrest, W. J., et al.  2004, \apjs,
154, 443
\bibitem[Furlan et al.(2005)]{furlan05} Furlan, E., Calvet, N., D'Alessio, P., 
Hartmann, L., Forrest, W. J., Watson, D. M., Uchida, K. I., Sargent, B., et al.
2005, \apjl, 628, L65
\bibitem[Furlan et al.(2006)]{furlan06} Furlan, E., Hartmann, L., Calvet, N., 
D'Alessio, P., Franco-Hern{\'a}ndez, R., Forrest, W. J., Watson, D. M., Uchida, K. I., 
et al. 2006, \apjs, 165, 568
\bibitem[Gauvin \& Strom(1992)]{gauvin92} Gauvin, L. S., \& Strom, K. M.  1992,
\apj, 385, 217
\bibitem[Geers et al.(2006)]{geers06} Geers, V. C., Augereau, J.-C., Pontoppidan,
K. M., Dullemond, C. P., Visser, R., Kessler-Silacci, J. E., Evans, N. J., van Dishoeck,
E. F., et al.  2006, \aap, 459, 545 
\bibitem[Geers et al.(2007)]{geers07} Geers, V. C., Pontoppidan, K. M., van Dishoeck,
E. F., Dullemond, C. P., Augereau, J.-C., Mer{\'\i}n, B., Oliveira, I., \& Pel, J. W.
2007, \aap, 469, L35
\bibitem[Ghez et al.(1993)]{ghez93} Ghez, A. M., Neugebauer, G., \& Matthews, K.
1993, \aj, 106, 2005
\bibitem[Ghez et al.(1995)]{ghez95} Ghez, A. M., Weinberger, A. J., Neugebauer, G.,
Matthews, K., \& McCarthy, D. W.  1995, \aj, 110, 753
\bibitem[Ghez et al.(1997)]{ghez97} Ghez, A. M., McCarthy, D. W., Patience, J. L.,
\& Beck, T. L.  1997, \apj, 481, 378
\bibitem[Goldreich \& Ward(1973)]{goldreich73} Goldreich, P., \& Ward, W. R.
1973, \apj, 183, 1051
\bibitem[G\'omez \& Mardones(2003)]{gomez03} G\'omez, M., \& Mardones, D.
2003, \aj, 125, 2134
\bibitem[Greene \& Young(1992)]{greene92} Greene, Th. P., \& Young, E. T.  1992,
\apj, 395, 516
\bibitem[Greene et al.(1994)]{greene94} Greene, Th. P., Wilking, B. A., Andr\'e, P.,
Young, E. T., \& Lada, Ch. J. 1994, \apj, 434, 614
\bibitem[Guenther et al.(2007)]{guenther07} Guenther, E. W., Esposito, M., Mundt, R.,
Covino, E., Alcal\'a, J. M., Cusano, F., \& Stecklum, B.  2007, \aap, 467, 1147
\bibitem[Haisch et al.(2001)]{haisch01} Haisch, K. E., Lada,  E. A., \& Lada, Ch. J. 
2001, \apjl, 553, L153
\bibitem[Harker \& Desch(2002)]{harker02} Harker, D. E., \& Desch, S. J.
2002, \apjl, 565, L109
\bibitem[Hartigan et al.(1994)]{hartigan94} Hartigan, P., Strom, K. M., \& Strom, 
S. E.  1994, \apj, 427, 961
\bibitem[Hartigan \& Kenyon(2003)]{hartigan03} Hartigan, P., \& Kenyon, S. J.
2003, \apj, 583, 334
\bibitem[Hartmann(1998)]{hartmann98} Hartmann, L. 1998, Accretion Processes
in Star Formation (Cambridge: Cambridge Univ. Press)
\bibitem[Hartmann et al.(1998)]{hartmann98b} Hartmann, L., Calvet, N., Gullbring, E.,
\& D'Alessio, P.  1998, \apj, 495, 385
\bibitem[Hartmann(2001)]{hartmann01} Hartmann, L.  2001, \aj, 121, 1030
\bibitem[Herbig \& Bell(1988)]{herbig88} Herbig, G. H., \& Bell, K. R.
1988, Lick Observatory Bulletin, Santa Cruz: Lick Observatory, 1988
\bibitem[Hern{\'a}ndez et al.(2005)]{hernandez05} Hern{\'a}ndez, J., Calvet, N.,
Hartmann, L., Brice\~no, C., Sicilia-Aguilar, A., \& Berlind, P.  2005, \aj, 129, 856
\bibitem[Hern{\'a}ndez et al.(2006)]{hernandez06} Hern{\'a}ndez, J., Brice\~no,
C., Calvet, N., Hartmann, L., Muzerolle, J., \& Quintero, A.  2006, \apj, 652, 472
\bibitem[Higdon et al.(2004)]{higdon04} Higdon, S. J. U., et al.  2004, \pasp,
116, 975
\bibitem[Hillenbrand et al.(1992)]{hillenbrand92} Hillenbrand, L. A., Strom, S. E.,
Vrba, F., \& Keene J.  1992, \apj, 397, 613
\bibitem[Houck et al.(2004)]{houck04} Houck, J. R., et al.  2004, \apjs, 154, 18
\bibitem[Hughes et al.(2009)]{hughes09} Hughes, A. M., Andrews, S. M., 
Espaillat, C., Wilner, D. J., Calvet, N., D'Alessio, P., Qi, C., Williams, J. P., \&
Hogerheijde, M. R.  2009, \apj, 698, 131
\bibitem[Ichikawa \& Nishida(1989)]{ichikawa89} Ichikawa, T., \& Nishida, M. 1989,
\aj, 97, 1074
\bibitem[Ireland \& Kraus(2008)]{ireland08} Ireland, M. J., \& Kraus, A. L.  2008,
\apjl, 678, L59
\bibitem[J\"ager et al.(1998)]{jaeger98} J\"ager, C., Molster, F. J., Dorschner, J.,
Henning, Th., Mutschke, H., Waters, L. B. F. M.  1998, \aap, 339, 904
\bibitem[Jayawardhana et al.(1999)]{jayawardhana99} Jayawardhana, R.,
Hartmann, L., Fazio, G., Fisher, R. S., Telesco, Ch. M., \& Pi\~na, R. K.  1999, 
\apjl, 521, L129
\bibitem[Jayawardhana et al.(2006)]{jayawardhana06} Jayawardhana, R., Coffey,
J., Scholz, A., Brandeker, A., \& van Kerkwijk, M. H.  2006, \apj, 648, 1206
\bibitem[Jensen et al.(2004)]{jensen04} Jensen, E. L. N., Mathieu, R. D., Donar,
A. X., \& Dullighan, A.  2004, \apj, 600, 789 
\bibitem[Kenyon et al.(1990)]{kenyon90} Kenyon, S. J., Hartmann, L. W., 
Strom, K. M., \& Strom, S. E.  1990, \aj, 99, 869
\bibitem[Kenyon \& Hartmann(1995)]{kenyon95} Kenyon, S. J., \& Hartmann, L.
1995, \apjs, 101, 117
\bibitem[Kenyon et al.(1998)]{kenyon98} Kenyon, S. J., Brown, D. I., Tout, Ch. A., 
\& Berlind, P.  1998, \aj, 115, 2491
\bibitem[Kessler-Silacci et al.(2006)]{kessler06} Kessler-Silacci, J., Augereau, J.-C., 
Dullemond, C. P., Geers, V., Lahuis, F., Evans, N. J., van Dishoeck, E. F., Blake, G. A.,
et al. 2006, \apj, 639, 275 
\bibitem[Kim et al.(2009)]{kim09} Kim, K. H., Watson, D. M., Manoj, P., Furlan, E.,
Najita, J., Forrest, W. J., Sargent, B., Espaillat, C., et al. 2009, \apj, 700, 1017
\bibitem[Lada(1987)]{lada87} Lada, Ch. J. 1987, in Star Forming Regions, 
proceedings of the IAU Symposium No. 115, ed. M. Peimbert \& J. Jugaku, 
Dordrecht:Reidel, 1
\bibitem[Lafreni\`ere et al.(2008)]{lafreniere08} Lafreni\`ere, D., Jayawardhana, R., 
Brandeker, A., Ahmic, M., \& van Kerkwijck, M. H.  2008, \apj, 683, 844
\bibitem[Leinert et al.(1993)]{leinert93} Leinert, Ch., Zinnecker, H., Weitzel, N.,
Christou, J., Ridgway, S. T., Jameson, R., Haas, M., \& Lenzen, R. 1993, \aap, 
278, 129
\bibitem[Lin \& Papaloizou(1986)]{lin86} Lin, D. N. C., \& Papaloizou, J.  1986,
\apj, 309, 846
\bibitem[Lin et al.(2000)]{lin00} Lin, D. N. C., Papaloizou, J. C. B., Terquem, C., 
Bryden, G., \& Ida, S.  2000, in Protostars and Planets IV, eds. V. Mannings, A. P. Boss,
\& S. S. Russell (Tucson: Univ. Arizona Press), 1111
\bibitem[Loinard et al.(2008)]{loinard08} Loinard, L., Torres, R. M., Mioduszewski, A.,
\& Rodr{\'\i}guez, L. F. 2008, \apjl, 675, L29
\bibitem[Loren(1989)]{loren89} Loren, R. B.  1989, \apj, 338, 902
\bibitem[Luhman(2000)]{luhman00} Luhman, K. L.  2000, \apj, 544, 1044
\bibitem[Luhman(2004a)]{luhman04a} Luhman, K. L.  2004a, \apj, 602, 816
\bibitem[Luhman(2004b)]{luhman04b} Luhman, K. L.  2004b, \apj, 617, 1216
\bibitem[Luhman(2007)]{luhman07} Luhman, K. L.  2007, \apjs, 173, 104
\bibitem[Luhman \& Rieke(1999)]{luhman99} Luhman, K. L., \& Rieke, G. H.
1999, \apj, 525, 440
\bibitem[Luhman et al.(2003)]{luhman03} Luhman, K. L., Stauffer, J. R., 
Muench, A. A., Rieke, G. H., Lada, E. A., Bouvier, J., \& Lada, C. J.  2003,
\apj, 593, 1093
\bibitem[Luhman et al.(2009)]{luhman09} Luhman, K. L., Allen, P. R., Espaillat, C.,
Hartmann, L., \& Calvet, N.  2009, \apj, submitted
\bibitem[Marsh \& Mahoney(1992)]{marsh92} Marsh, K. A., \& Mahoney, M. J.
1992, \apj, 395, L115
\bibitem[Mart{\'\i}n et al.(1998)]{martin98} Mart{\'\i}n, E. L., Montmerle, T.,
Gregorio-Hetem, J., \& Casanova, S.  1998, \mnras, 300, 733
\bibitem[Mathis(1990)]{mathis90} Mathis, J. S. 1990, \araa, 28, 37
\bibitem[McClure et al.(2008)]{mcclure08} McClure, M. K., Forrest, W. J., 
Sargent, B. A., Watson, D. M., Furlan, E., Manoj, P., Luhman, K. L., Calvet, N.,
et al.  2008, \apjl, 683, L187
\bibitem[McClure(2009)]{mcclure09} McClure, M. K. 2009, \apjl, 639, L81
\bibitem[Meeus et al.(2003)]{meeus03} Meeus, G., Sterzik, M., Bouwman, J., \&
Natta, A.  2003, \aap, 409, L25
\bibitem[Melo(2003)]{melo03} Melo, C. H. F.  2003, \aap, 410, 269
\bibitem[Meyer et al.(1997)]{meyer97} Meyer, M. R., Calvet, N., \& Hillenbrand, 
L. A.  1997, \aj, 114, 288
\bibitem[Miyake \& Nakagawa(1995)]{miyake95} Miyake, K., \& Nakagawa, Y.
1995, \apj, 441, 361
\bibitem[Natta et al.(2006)]{natta06} Natta, A., Testi, L., \& Randich, S.  2006, 
\aap, 452, 245
\bibitem[Nelson et al.(2000)]{nelson00} Nelson, R. P., Papaloizou, J. C. B., Masset,
F., \& Kley, W.  2000, \mnras, 318, 18
\bibitem[Padgett et al.(2008)]{padgett08} Padgett, D. L., Rebull, L. M., Stapelfeldt,
K. R., Chapman, N. L., Lai, S.-P., Mundy, L. G., Evans, N. J., Brooke, T. Y., et al.
2008, \apj, 672, 1013
\bibitem[Palla \& Stahler(2000)]{palla00} Palla, F., \& Stahler, S. W.  2000,
\apj, 540, 255
\bibitem[Persi et al.(2000)]{persi00} Persi, P., Marenzi, A. R., Olofsson,G., Kaas, 
A. A., Nordh, L., Huldtgren, M., Abergel, A., Andr{\'e}, P., et al.  2000, \aap, 
357, 219 
\bibitem[Pi{\'e}tu et al.(2006)]{pietu06} Pi{\'e}tu, V., Dutrey, A., Guilloteau, S.,
Chapillon, E., \& Pety, J.  2006, \aap, 460, L43
\bibitem[Prato et al.(2003)]{prato03} Prato, L., Greene, T. P., \& Simon, M.  
2003, \apj, 584, 853
\bibitem[Preibisch et al.(1998)]{preibisch98} Preibisch, Th., Guenther, E., Zinnecker,
H., Sterzik, M., Frink, S., \& R{\"o}ser, S.  1998, \aap, 333, 619
\bibitem[Przygodda et al.(2003)]{przygodda03} Przygodda, F., van Boekel, R.,
Abraham, P., Melnikov, S. Y., Waters, L. B. F. M., \& Leinert, Ch.  2003, \aap,
412, L43
\bibitem[Quillen et al.(2004)]{quillen04} Quillen, A. C., Blackman, E. G., Frank, A.,
\& Varni\`ere, P.  2004, \apjl, 612, L137
\bibitem[Ratzka et al.(2005)]{ratzka05} Ratzka, T., K\"ohler, R., \& Leinert, Ch.  
2005, \aap, 437, 611
\bibitem[Reipurth \& Zinnecker(1993)]{reipurth93} Reipurth, B., \& Zinnecker, H.
1993, \aap, 278, 81
\bibitem[Sargent et al.(2006)]{sargent06} Sargent, B., Forrest, W. J., D'Alessio, P.,
Li, A., Najita, J., Watson, D. M., Calvet, N., Furlan, E., et al. 2006, \apj, 645, 395
\bibitem[Sargent et al.(2009)]{sargent09} Sargent, B. A., Forrest, W. J., Tayrien, C.,
McClure, M. K., Watson, D. M., Sloan, G. C., Li, A., Manoj, P., et al.  2009, \apjs,
182, 477
\bibitem[Schegerer et al.(2008)]{schegerer08} Schegerer A. A., Wolf, S., Ratzka, Th.,
\& Leinert, Ch.  2008, \aap, 478, 779
\bibitem[Sicilia-Aguilar et al.(2006)]{sicilia06} Sicilia-Aguilar, A., Hartmann, L., Calvet, N., 
Megeath, S. T., Muzerolle, J., Allen, L., D'Alessio, P., Mer{\'\i}n, B., et al. 2006, 
\apj, 638, 897
\bibitem[Simon et al.(1995)]{simon95} Simon, M., Ghez, A. M., Leinert, Ch., 
Cassar, L., Chen, W. P., Howell, R. R., Jameson, R. F., Matthews, K., Neugebauer, G., 
\& Richichi, A.  1995, \apj, 443, 625
\bibitem[Skrutskie et al.(1990)]{skrutskie90} Skrutskie, M. F., Dutkevitch, D., 
Strom, S. E., Edwards, S., \& Strom, K. M. 1990, \aj, 99, 1187 
\bibitem[Skrutskie et al.(2006)]{skrutskie06} Skrutskie, M. F., et al.  2006, \aj,
131, 1163
\bibitem[Strom \& Strom(1994)]{strom94} Strom, K. M., \& Strom, S. E.  1994,
\apj, 424, 237
\bibitem[Takeuchi et al.(1996)]{takeuchi96} Takeuchi, T., Miyama, S. M., \&
Lin, D. N. C.  1996, \apj, 460, 832
\bibitem[Tanaka \& Ida(1999)]{tanaka99} Tanaka, H., \& Ida, S.  1999, Icarus,
139, 350
\bibitem[Torres et al.(1995)]{torres95} Torres, C. A. O., Quast, G., de la Reza, R.,
Gregorio-Hetem, J., \& L\'epine, J. R. D.  1995, \aj, 109, 2146
\bibitem[Torres et al.(2006)]{torres06} Torres, C. A. O., Quast, G. R., da Silva, L.,
de la Reza, R., Melo, C. H. F., \& Sterzik, M.  2006, \aap, 460, 695
\bibitem[Torres et al.(2007)]{torres07} Torres, R. M., Loinard, L., Mioduszewski,
A. J., \& Rodr{\'\i}guez, L. F.  2007, \apj, 671, 1813
\bibitem[van Boekel et al.(2003)]{vanboekel03} van Boekel, R., Waters, L. B. F. M., 
Dominik, C., Bouwman, J., de Koter, A., Dullemond, C. P., \& Paresce, F.  2003,
\aap, 400, L21
\bibitem[Varni\`ere et al.(2006)]{varniere06} Varni\`ere, P., Blackman, E. G., 
Frank, A., \& Quillen, A. C.  2006, \apj, 640, 1110
\bibitem[Vieira et al.(2003)]{vieira03} Vieira, S. L. A., Corradi, W. J. B., Alencar,
S. H. P., Mendes, L. T. S., Torres, C. A. O., Quast, G. R., Guimaraes, M. M., \&
da Silva, L.  2003, \aj, 126, 2971
\bibitem[Watson et al.(2009)]{watson09} Watson, D. M., Leisenring, J. M., Furlan, E.,
Bohac, C. J., Sargent, B., Forrest, W. J., Calvet, N., Hartmann, L., et al.  2009, \apjs,
180, 84
\bibitem[Wenrich \& Christensen(1996)]{wenrich96} Wenrich, M. L., \& Christensen, 
P. R. 1996, \jgr, 101, 15921
\bibitem[Werner et al.(2004)]{werner04} Werner, M. W. et al.  2004, \apjs, 
154, 1
\bibitem[White \& Basri(2003)]{white03} White, R. J., \& Basri, G.  2003, \apj,
582, 1109
\bibitem[White \& Ghez(2001)]{white01} White, R. J., \& Ghez, A. M.  2001,
\apj, 556, 265
\bibitem[White \& Hillenbrand(2004)]{white04} White, R. J., \& Hillenbrand, L. A.
2004, \apj, 616, 998
\bibitem[Whittet et al.(1997)]{whittet97} Whittet, D. C. B., Prusti, T., Franco, G. A. P.,
Gerakines, P. A., Kilkenny, D., Larson, K. A., \& Wesselius, P. R.  1997, \aap, 327, 1194
\bibitem[Whittet et al.(1988)]{whittet88} Whittet, D. C. B., Bode, M. F., Longmore,
A. J., Adamson, A. J., McFadzean, A. D., Aitken, D. K., \& Roche, P. F. 1988,
\mnras, 233, 321
\bibitem[Whittet et al.(2004)]{whittet04} Whittet, D. C. B., Shenoy, S. S., Clayton, 
G. C., \& Gordon, K. D.  2004, \apj, 602, 291
\bibitem[Wilking et al.(1989)]{wilking89} Wilking, B. A., Lada, Ch. J., \& Young, E. T.  
1989, \apj, 340, 823
\bibitem[Wilking et al.(2005)]{wilking05} Wilking, B. A, Meyer, M. R., Robinson,
J. G., \& Greene, Th. P.  2005, \aj, 130, 1733
\end{thebibliography}
\end{document}